\documentclass[journal]{IEEEtran}

\usepackage{pifont}
\usepackage{cite}
\usepackage[hidelinks]{hyperref}
\usepackage{url}
\usepackage{multirow}
\usepackage{algorithm}
\usepackage{algorithmic}
\usepackage{color}
\usepackage{amssymb}
\usepackage{subcaption}
\usepackage[font=scriptsize,labelfont=bf]{caption}

\def\Min{{\rm minimize }}
\def\Max{{\rm maximize }}
\def\ST{{\rm subject\ to}}

\def\b0{{\bf 0}}
\def\bb1{{\bf 1}}

\ifCLASSINFOpdf
   \usepackage[pdftex]{graphicx}
   \usepackage[table]{xcolor}
   
   \graphicspath{{../pdf/}{figures/}}

   \DeclareGraphicsExtensions{.pdf,.jpeg,.png,.jpg}
\else

\fi

\usepackage[cmex10]{amsmath}

\usepackage{algorithmic}
\usepackage{algorithm}
\usepackage{array}
\newcolumntype{L}[1]{>{\raggedright\let\newline\\\arraybackslash\hspace{0pt}}m{#1}}
\newcolumntype{C}[1]{>{\centering\let\newline\\\arraybackslash\hspace{0pt}}m{#1}}
\newcolumntype{R}[1]{>{\raggedleft\let\newline\\\arraybackslash\hspace{0pt}}m{#1}}

\usepackage[cmex10]{amsmath}
\usepackage{lineno,xcolor}

\usepackage[printonlyused]{acronym}

\acrodef{ILP}{Integer Linear Program}
\acrodef{ISG}{Industry Standards Group}
\acrodef{ETSI}{European Telecommunications Standards Institute}
\acrodef{InP}{Infrastructure Provider}
\acrodef{VN}{Virtual Network}
\acrodef{SN}{Substrate Network}
\acrodef{VNE}{Virtual Network Embedding}
\acrodef{SP}{Service Provider}
\acrodef{TSP}{Telecommunication Service Provider}
\acrodef{NFV}{Network Function Virtualization}
\acrodefplural{CAPEX}[CAPEX]{CAPital EXpenses}
\acrodefplural{OPEX}[OPEX]{OPerating EXpenses}

\makeatletter
\def\ps@IEEEtitlepagestyle{%
  \def\@oddhead{\strut\hfill 
  
\begin{tabular}{m{17.5cm}}
          \centering
          \footnotesize{This is the author's version of an article that has been published in IEEE Transactions on Network and Service Management. Changes were made to this version by the publisher prior to publication. The final version of record is available at {\color{blue}http://dx.doi.org/10.1109/TNSM.2015.2459073}}
          \tabularnewline
          \end{tabular}%
\hfill\strut}%
  \def\@oddfoot{\mycopyrightnotice}%
  \def\@evenfoot{}%
}

\def\mycopyrightnotice{%
  {
\begin{tabular}{m{19.0cm}}
          \centering
          \footnotesize{Copyright (c) 2015 IEEE. Personal use is permitted. For any other purposes, permission must be obtained from the IEEE by emailing pubs-permissions@ieee.org.}
          \tabularnewline
          \end{tabular}%
  
  \hfill}
  \gdef\mycopyrightnotice{}
}

\begin{document}

\title{A Path Generation Approach to Embedding of Virtual Networks}

\author{Rashid~Mijumbi, Joan~Serrat, Juan-Luis~Gorricho, Raouf~Boutaba

\thanks{R. Mijumbi, J. Serrat and J.L. Gorricho are with the Network Engineering Department, Universitat Polit\`{e}cnica de Catalunya, 08034 Barcelona, Spain.}
\thanks{R. Boutaba is with the D.R. Cheriton School of Computer Science, University of Waterloo, Waterloo, Ontario, N2L 3G1, Canada.}

}

\markboth{IEEE TRANSACTIONS ON NETWORK AND SERVICE MANAGEMENT}
{Shell \MakeLowercase{\textit{et al.}}: Bare Demo of IEEEtran.cls for Journals}
%



\maketitle

\begin{abstract}
As the virtualization of networks continues to attract attention from both industry and academia, the \ac{VNE} problem remains a focus of researchers. This paper proposes a one-shot, unsplittable flow \ac{VNE} solution based on column generation. We start by formulating the problem as a path-based mathematical program called the primal, for which we derive the corresponding dual problem. We then propose an initial solution which is used, first, by the dual problem and then by the primal problem to obtain a final solution. Unlike most approaches, our focus is not only on embedding accuracy but also on the scalability of the solution. In particular, the one-shot nature of our formulation ensures embedding accuracy, while the use of column generation is aimed at enhancing the computation time to make the approach more scalable. In order to assess the performance of the proposed solution, we compare it against four state of the art approaches as well as the optimal link-based formulation of the one-shot embedding problem. Experiments on a large mix of \ac{VN} requests show that our solution is near optimal (achieving about 95\% of the acceptance ratio of the optimal solution), with a clear improvement over existing approaches in terms of \ac{VN} acceptance ratio and average \ac{SN} resource utilization, and a considerable improvement (92\% for a \ac{SN} of 50 nodes) in time complexity compared to the optimal solution.

\end{abstract}


 \ifCLASSOPTIONpeerreview
 \begin{center} \bfseries EDICS Category: 3-BBND \end{center}
 \fi

\IEEEpeerreviewmaketitle

\section{Introduction}

The ever increasing requirements placed on the Internet are fueling its evolution to architectures which make a better and more efficient use of the available network resources, and promote service innovations. \acp{SP} have to satisfy personalized needs for their customers and hence they are impelled to use different protocol stacks and provide customized services and network resources. Network virtualization \cite{Fischer13} has been proposed as a feasible solution to achieve this goal. In network virtualization, \acp{InP} divide their resources into chunks, called \acp{VN}, which are allocated to \acp{SP}. Thanks to virtualization, the resource chunks are isolated from each other so the service networks behave as if they were independent though they share the same substrate infrastructure.\\
\indent However, the creation of \acp{VN} on top of a \ac{SN} is not trivial. \ac{VN} topologies composed of virtual nodes and virtual links have to be drawn to support traffic flows from sources to sinks. Virtual nodes and virtual links then have to be mapped onto the physical substrate in a way that satisfies user demands and optimizes the use of the available resources. This is the basis of the so called \ac{VNE} problem \cite{Fischer13}, which in case of unsplittable flows, i.e. flows that have to be treated as a unit from source to sink, is NP hard \cite{Chowdhury12}. Therefore, to simplify the problem, several existing solutions to VNE either assume that the \ac{SN} supports the splitting of flows \cite{Yu08}, or carry out the node and link embedding in two separate steps \cite{Zhu06}, which can lead to blocking or rejecting of resource requests at the link mapping stage and hence a sub-optimal substrate resource utilization.\\
\indent In this paper, we propose a near optimal solution to the unsplittable flow VNE problem obtained by performing the embedding in one-shot (i.e. both virtual nodes and links are embedded in one step) using mathematical programming and path generation\footnote{In this paper, we use the terms \emph{path generation} and \emph{column generation} interchangeably.} \cite{Barnhart98}. The formulation of the embedding problem as being one-shot is motivated by the need to obtain an efficient embedding solution (which would ultimately lead to better resource utilization and hence better profitability for \acp{InP}), while the employment of path generation is aimed at ensuring that the resulting algorithm is more scalable compared to the optimal formulation.\\
\indent To this end, we formulate two mathematical programs; one is a path-based formulation of the unsplittable flow one-shot VNE problem, also known as the primal problem, while the other is its corresponding dual problem. For given instances of the problem, both the primal and dual problems have approximately the same solution value. The proposed approach begins by obtaining an initial solution (composed of paths in an augmented \ac{SN}) to the primal problem using a VNE approach that performs node and link mapping in two coordinated stages. The next step is to enhance the initial solution. This is achieved by using the initial solution as an input into the dual problem, hence resulting into \textit{prices} for the \ac{SN} links and nodes. Using  Dijkstra's algorithm \cite{Cormen09}, these prices are utilized to determine an additional set of paths which can be added to enhance the solution. These paths, together with those obtained in the initial solution, are finally used to solve the primal problem to obtain a final embedding solution.

The main contributions of this paper are as follows:
\begin{itemize}
  \item A near optimal unsplittable flow one-shot \ac{VNE} approach that improves substrate resource utilization compared to existing heuristic and approximation solutions.
  \item A path generation-based approach for unsplittable flows that significantly improves the time complexity of the embedding compared to the optimal solution.
\end{itemize}
The rest of the paper is organized as follows: Section II presents the description of the \ac{VNE} problem. We present the related work in Section III. Sections IV  and V respectively describe the mathematical formulation of the one-shot embedding problem and its solution based on path generation. Section IV presents the evaluation of our proposed solution and the discussion of the results. Finally, Section VII concludes this paper.
\section{Problem Formulation}
\subsection{Substrate Network Capacity}
We model the \ac{SN} as an undirected graph denoted by $G_s = (N_s, L_s)$, where $N_s$ and $L_s$ represent the set of substrate nodes and links, respectively. Each substrate link $l_{uv}\in L_s$ connecting the nodes $u$ and $v$ has a bandwidth capacity $C_{uv}$ while each substrate node $u\in N_s$ has computation capacity $C_u$ and a location $Loc_u(x,y)$
\subsection{Virtual Network Requests}
In the same way, we model the \ac{VN} as an undirected graph denoted by $G_v = (N_v, L_v)$, where $N_v$ and $L_v$ represent the set of virtual nodes and links respectively. Each virtual link $l_{ij}\in L_v$ connecting the nodes $i$ and $j$ has a bandwidth demand $D_{ij}$ while each virtual node $i\in N_v$ has computation demand $D_i$, a location $Loc_i(x,y)$ as well a constraint on its location $Dev_i(\Delta x, \Delta y)$ which specifies the maximum allowed deviation for each of its $x$ and $y$ coordinates\footnote{The notation used in this paper is to represent virtual nodes with the letters $i$ or $j$ and substrate nodes with $u$ or $v$.}. Constraints on the location of virtual nodes are aimed at giving \acp{SP} the flexibility to choose the geographical placement of given parts of their network topologies. This could be as a result of a given \ac{SP} introducing specialized services for users in a given location, or a desire to ensure improved quality of service by restricting the distance (and hence latency) between a given pair of nodes.

\subsection{Virtual Network Embedding}
The embedding problem consists in the mapping of each virtual node $i \in N_v$ to one of the possible substrate nodes with in the set $\Upsilon(i)$. $\Upsilon(i)$ is defined as a set of all substrate nodes $u \in N_s$ which have enough \emph{available capacity} (defined the difference between the total capacity of a resource and the amount already allocated) and are \emph{located} within the maximum allowed deviation $Dev_{i}(\Delta x, \Delta y)$ of the virtual node. For a successful mapping, each virtual node must be mapped and any given substrate node can only map at most one virtual node from the same request.
Similarly, all the virtual links have to be mapped to one or more substrate links connecting the nodes to which the virtual nodes at its ends have been mapped. Each of the substrate links must have enough capacity to support the virtual link(s) that go through it. A mapping is successful if all the virtual links are mapped.

\begin{figure}[t!]
 \centering
 \hspace*{-0.25in}
  \includegraphics[width=0.40\paperwidth,height=0.20\paperheight]{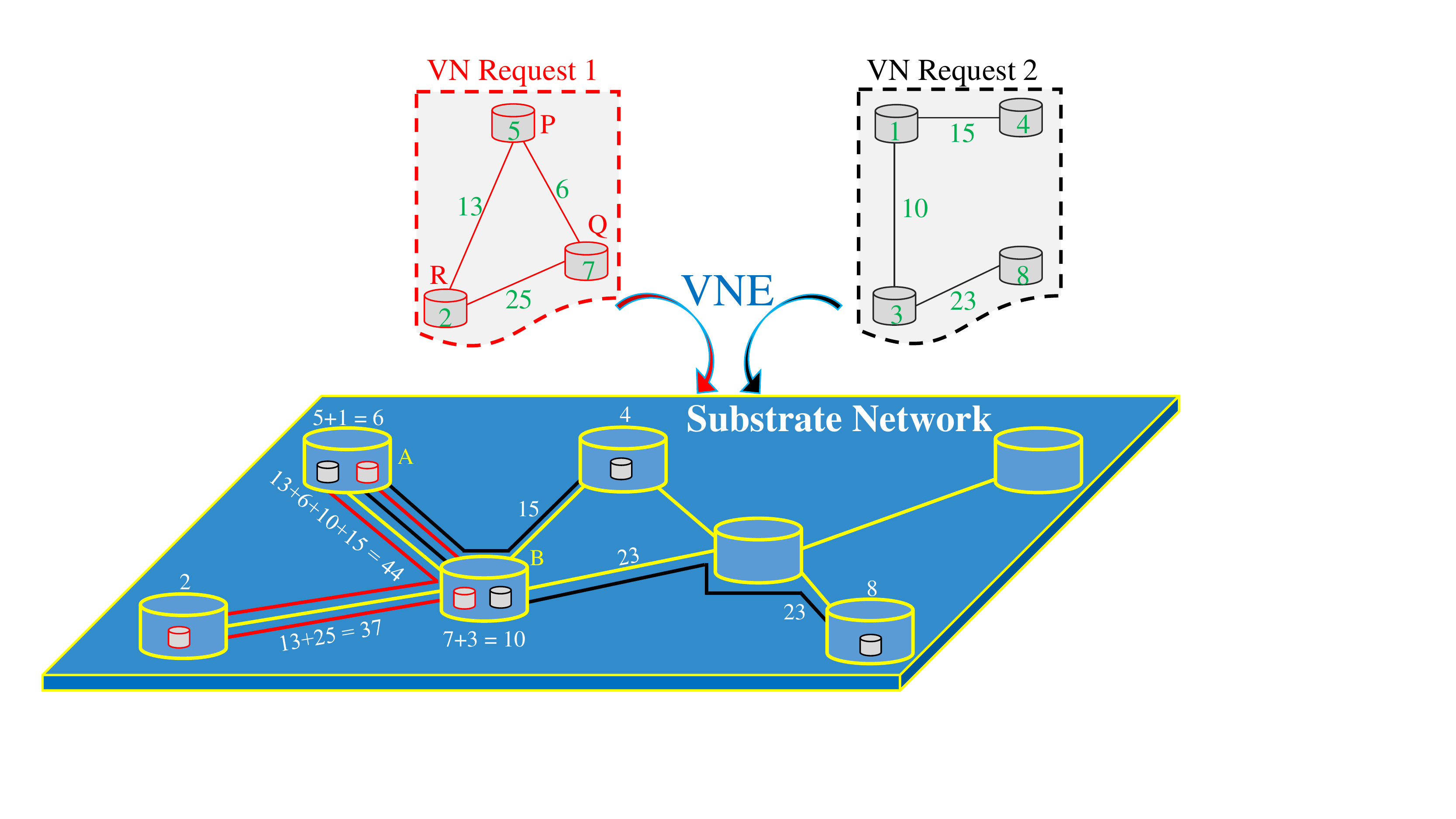}\\
  \caption{Virtual Network Embedding: Two VNs mapped onto a SN}
  \label{vne}
\end{figure}

In Fig. \ref{vne}, we show an example of two \acp{VN} being mapped onto a \ac{SN}. The resource requirements for each virtual node or link is also shown. The values in the \ac{SN} are the total loading of any given physical node or link. As can be noted from Fig. \ref{vne}, one substrate node can host more than one virtual node (e.g. node A). A substrate link can also host more than one virtual link (e.g. link AB), and a given virtual link can span more than one substrate link (e.g link RP).

In general, the objective in \ac{VNE} is to map as many \acp{VN} as possible, hence leading to an efficient utilization of \ac{SN} resources. For the online \ac{VNE} problem, there is no knowledge on the requirements of future \ac{VN} requests, and as such, one way of ensuring that as many requests are accepted is by balancing the overall loading of the \ac{SN} \cite{Chowdhury12} such that all substrate resources (nodes and links) are equally likely to accept resource requests. In the same way, it is worth noting that due to the lack of information about future virtual network requests, optimality as referred to in this paper is only based on the mapping of an arriving virtual network request to a substrate network, which could possibly already have other virtual networks already embedded, or for which other virtual networks may be embedded in the future. Therefore, this optimality does not represent an optimal embedding solution considering all possible virtual network requests.

\section{Related Work}
\ac{VNE} is a well-studied problem. In what follows, we only discuss those approaches we consider more closely related to our proposal. An interested reader is referred to \cite{Fischer13} for a detailed survey on \ac{VNE}.
\subsection{Two-step Embedding}
Some approaches based on two stages, starting with node mapping and then link mapping, are proposed in \cite{Yu08} and \cite{Lu06}. These algorithms measure the resource of a node or link by its CPU capacity, or bandwidth without considering the topological structure of the VNs and the underlying substrate network. However, the topological attributes of nodes may have an impact on the success and efficiency of VNE. Cheng et al. \cite{ChengXiang2011} propose a topology-aware node mapping approach which uses the Markov Random Walk model to rank virtual and substrate network nodes based on their resource and topological attributes. The links are then mapped either using the shortest path (for unsplittable flows), or formulated as a commodity flow problem for splittable flows. Unlike our work, the above approaches don't consider location constraints on virtual nodes, assuming that they can be mapped at any location in the \ac{SN}. A coordinated node and link mapping is proposed in \cite{Chowdhury12}. Although the coordination here improves the solution space, the mapping is still performed in two separate stages, hence yielding sub-optimal embedding.

The works in \cite{He08, mijumbi1, mijumbi2, mijumbi3} propose dynamic and distributed approaches to \ac{VN} resource allocation, where the actual resources allocated to virtual nodes and links is scaled up and down based on actual resources utilization as well as resource availability. However, they do not consider the embedding stage, assuming that the \ac{VN} is already mapped to a \ac{SN}.

\subsection{One-shot Embedding}
A one-shot embedding solution based on a multi-agent system is proposed in \cite{Houidi08}. However, this proposal assumes unbounded \ac{SN} resources, and all \ac{VN} requests to be known in advance. Also, messaging overhead exchanged between the agents can be detrimental to solution scalability. Zhu et al. \cite{Zhu06} also propose a one-shot mapping solution, assuming infinite substrate resources, and no constraints on the locations of nodes. Authors in \cite{Infuhr11} - \cite{Schaffrath10} propose different approaches to one-shot VNE assuming that all VN requests are known in advance (offline solutions), while those in \cite{Lischka09} - \cite{Yu10} make simplifying assumptions with regard to the capacity of the \ac{SN} and do not consider constraints on virtual nodes locations. While most VNE proposals use topologies to represent VN requests, \cite{Wang11} proposes the use of traffic matrices. However, the embedding is achieved by alleviating constraints on \ac{VN} resources, such as node location. Houidi et al. \cite{Houidi11} split any given VN request across multiple infrastructure providers and then uses max-flow and min-cut algorithms and linear programming to find one-shot solutions to the partial VN graphs. While the embeddings of the split graphs are solved in one-shot, they do not encompass the original VN request in its entirety.

Perhaps the the works most related our work are by Jarray et al. \cite{JarrayA14} and Hu et al. \cite{QianHUYang13} both of whom apply column generation to VNE. Jarray et al. apply a column generation approach to one-shot VNE by assuming that the embedding of \ac{VN} requests can be delayed by storing each arriving request to process them in batches using an auctioning mechanism. The proposal can therefore be considered to be an offline one. Hu et al. formulate a one-shot path-based VNE where the virtual links are represented as commodities. The formulated mathematical program is then relaxed so as to apply column generation. However, Hu et al. consider a scenario where the demand/commodity of any given virtual link may be split over more than one substrate path. This differs from the proposal in this paper which solves a harder problem where the flows are not splittable.

\subsection{Mathematical Programming}
Mathematical programming has been applied to a variety of problems in networking. Xie et al. \cite{Xie12} use mathematical programming for dynamic resource allocation in networks while Botero et al. \cite{Botero12} use an optimization technique for link mapping (assuming that the virtual nodes have already been mapped to substrate nodes). Unlike all these works, the mathematical programming formulation proposed in this paper does not only focus on unsplittable flows, but also combines both node and link mapping in one stage. The node mapping step is an important part of VNE since it determines the efficiency of the link mapping. This is why such approaches that coordinate these two steps have been shown to lead to better resource utilization efficiency \cite{Chowdhury12}. Combining these two steps together even further enhances this efficiency, yet the resulting mathematical program is even harder to solve. Finally, path generation based formulations for multi-commodity flow based problems are proposed in a number of approaches such as \cite{Santos10}. In these formulations the source and end nodes for each flow are known a priori, which reduces the complexity of the problem, compared to the one-shot \ac{VNE} that we solve in this paper.
\subsection{Summary}
To summarize, because of the NP hardness of the \ac{VNE} problem, existing one-shot approaches either make simplifying assumptions such as considering the offline version of the problem, assuming infinite resources, or ignoring constraints on the virtual nodes and links, while other proposals solve the embedding problem in two stages, typically employing a greedy approach for node mapping and then try to optimize the link mapping. The approach proposed in this paper differs from previous work in many aspects. Most applications of mathematical programming and path generation to routing are concerned with simpler problems, in which either both the source and sink nodes are known, in which case the problem reduces to a load balancing problem, or only consider node mapping. While the link mapping \emph{sub-problem} is still NP-hard for unsplittable flows, it is even harder in the case of \ac{VNE} since the source and sink nodes should also be determined. To the best of our knowledge, this is the first path-based mathematical programming solution to the one-shot, unsplittable flow \ac{VNE} problem.
\section{One-Shot Virtual Network Embedding}
The one-shot \ac{VNE} problem involves performing both node and link mapping at the same time. In this paper, we use mathematical programming to achieve this. Specifically, we consider that VNs arrive one at a time following a Poisson distribution and have exponentially distributed service times, and the formulated optimization problem involves the embedding of a single VN at any given time. This way, at every mapping step, the actual resource availability of all substrate links and nodes is taken into account when performing a mapping. For reference, the link-based formulation of the problem that obtains an optimal solution using mathematical programming is shown in the appendix. Here, we adopt a path-based formulation using column generation in order to solve the problem with much less time and storage requirements.
\subsection{Substrate Network Augmentation}
We start by creating an augmented network first introduced in \cite{Chowdhury12},  with each virtual node $i$ connected to each of the substrate nodes in its possible node set $\Upsilon(i)$ by a \emph{meta link} \cite{Chowdhury12} $l_{iu} \in L_x$, where $L_x$ is the set of all meta links. Then the aim is to establish a single path ${p_{uv}^{ij}}$ from each virtual node $i$ to all other virtual nodes $j$ to which it is connected. The path ${p_{uv}^{ij}}$ is made of two meta links, $l_{iu}$ and $l_{jv}$, and a sub-path in the \ac{SN} connecting the substrate nodes $u$ and $v$. This sub-path may be made up of one or more \ac{SN} links.
\begin{figure}
\centering
  {\includegraphics[width=8cm, height=5.5cm]{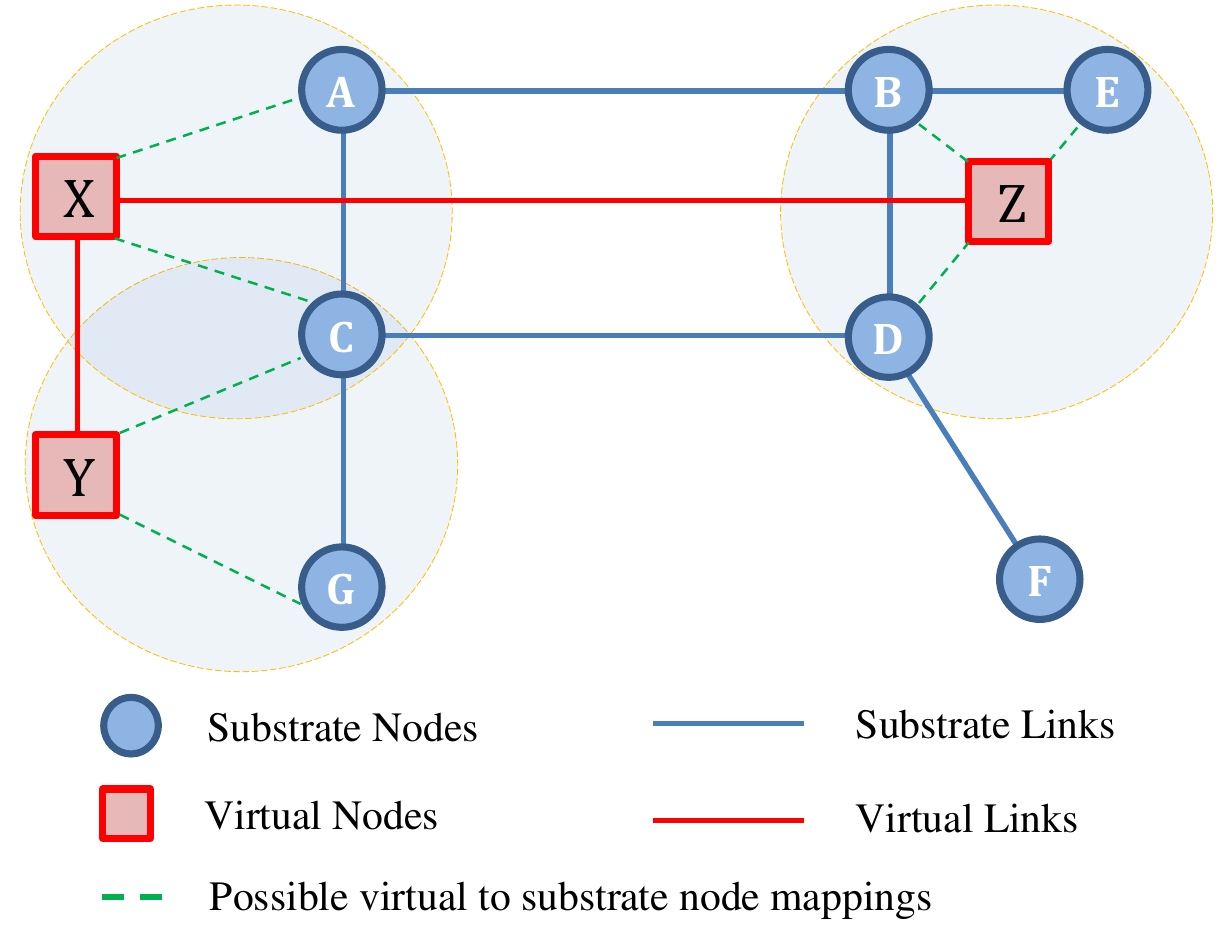}}
  \caption{VNE showing Virtual Node Mapping Constraints}
  \label{Diagram}
\end{figure}
In Fig. \ref{Diagram}, we show a representation of an instance of the problem. In the figure, XYZ are nodes of a \ac{VN}, while ABCDEFG are nodes of a \ac{SN}. As an example, for virtual link XZ, one possible path could be XABEZ, and is represented as ${p_{ae}^{xz}}$. The path ${p_{ae}^{xz}}$ is a sequence of links in the augmented network that start from one end of the virtual link to the other. Therefore, in order to embed the virtual link XZ, we need to determine the three components of the path, which $-$ for this example $-$ are the two meta links XA and EZ, and the \ac{SN} path ABE composed of two links, AB and BE. The components XA and EZ can be determined from a virtual to substrate node mapping, while ABE from a link mapping approach such as shortest path. In particular, this path example would mean that the virtual node X is mapped onto substrate node A, the virtual node Z is mapped onto substrate node E and that the virtual link XZ is mapped onto the \ac{SN} path ABE. One difficulty illustrated in this example comes from the fact that if, for example, we choose the path XABEZ for virtual link XZ, then the virtual link XY can only be mapped on a path that includes meta link XA and not XC. This would in turn require that Y be mapped onto C, otherwise we would have a sub-optimal solution in which the virtual link XY uses resources from two substrate links (AC \& CG) instead of a single link (CG). Hence, the determination of these paths should not be carried sequentially and independently. As previously mentioned, our aim is to find the best possible path for each of the virtual links subject to the mapping requirements described in our problem formulation (see Section II).
\subsection{\textbf{LP$-$P}: Path based Formulation $- Primal$}
We formulate the \ac{VNE} problem as a commodity flow problem\cite{Pfetsch06}, where virtual links are flows that should be carried by the \ac{SN}. However, unlike most commodity flow formulations, in our case, the source node $i$ and terminal node $j$ for each flow also need to be determined.\\\\
\textbf{Variable and Parameter definitions:} In this formulation, we define a non-negative binary variable $f_{uv}^{ij} = [0, D_{ij}]$ which represents the unsplittable flow of a virtual link $l_{ij} \in L_v$ on a simple substrate path $p_{uv}^{ij} \in P$. The indices $u$, $v$, $i$ and $j$ define a path $(i-u-v-j)$ in the augmented \ac{SN}. As described in IV (A), these paths are made up of three components:  two meta-links $iu$ and $jv$, and a \ac{SN} path from $u$ to $v$. The variable $f_{uv}^{ij}$ is binary in that it can only take on values $0$ and $D_{ij}$, where $D_{ij}$ is the demand of virtual link $l_{ij} \in L_v$. We define $P$ as a set of all the possible substrate paths, $P_{uv}$ as the set of all paths that use the substrate link $l_{uv} \in L_s$ and $P^{ij}$ as the set of all paths that can support the flow for virtual link $l_{ij} \in L_v$. We also define $\chi_{u}^{i}$ = [0,1] as a binary variable equal to 1 if the virtual node $i$ is mapped onto the substrate node $u$ and 0 otherwise. As mentioned in IV (A), it is important to note that variables $\chi_{u}^{i}$ and $\chi_{v}^{j}$ directly determine the existence or otherwise of meta links $iu$ and $jv$ for the path ${p_{uv}^{ij}}$ since the meta links are dependent on the respective node mappings. For example, if $\chi_{u}^{i} == 0$ then the virtual node $i$ is not mapped onto substrate node $u$, implying that the meta link from $i$ to $u$ is non existent, and so is the path ${p_{uv}^{ij}}$. Let $A_{uv}$ be the available bandwidth capacity on the substrate link $l_{uv}$, and $A_{u}$ be the available computation capacity on node $u$.\\\\
\textbf{Objective:} The objective of the mathematical formulation \eqref{start1}$-$\eqref{bigM2} is to balance the resource usage of the \ac{SN}, by favoring the selection of those resources with comparatively higher available capacity. Balancing the loading of the \ac{SN} has two advantages; first, it distributes the mapping of a given VN request over multiple \ac{SN} resources which avoids a single VN being majorly affected by single or regional failures in \ac{SN}, hence ensuring better VN survivability. In addition, since the problem we consider in this paper is online, we do not know in advance the required node locations for VN requests. Balancing the loading of the \ac{SN} ensures that at any given point, each substrate node/link has the same capacity on average. This avoids situations where a VN request would be rejected due to one or more of its nodes not being able to be mapped because substrate nodes in their respective possible node sets $\Upsilon(i)$ have less resources than other parts of the \ac{SN}. As was shown by \cite{Chowdhury12}, load balancing leads to a better acceptance ratio of \acp{VN}, which would directly translate in higher incomes for \acp{InP}.
\begin{equation}
\Min \sum_{l_{ij} \in L_v} \sum \limits_{p_{uv}^{ij} \in P}  \frac{1}{A_{uv}}f_{uv}^{ij} + \sum \limits_{i \in N_v} \sum \limits_{u \in \Upsilon(i)} \frac{1}{A_{u}}\chi_{u}^{i}
\label{start1}
\end{equation}
$\ST$
\begin{equation}
\sum \limits_{u \in \Upsilon(i)} \chi_{u}^{i} = 1 \hspace{10 mm}\forall i \in N_v
\label{vir1}
\end{equation}
\begin{equation}
\sum \limits_{i \in N_v} \chi_{u}^{i} \leq 1 \hspace{10 mm}\forall u \in N_s
\label{sub1}
\end{equation}
\begin{equation}
\sum \limits_{p_{uv}^{ij} \in P^{ij}} f_{uv}^{ij} = D_{ij}\hspace{10 mm}\forall l_{ij} \in L_v
\label{dem}
\end{equation}
\begin{equation}
\sum \limits_{p_{uv}^{ij} \in P_{uv}} f_{uv}^{ij}\leq A_{uv}\hspace{10 mm}\forall l_{uv} \in L_s
\label{cap}
\end{equation}
\begin{equation}
{f_{uv}^{ij}} - D_{ij} \chi_{u}^{i}\leq 0\hspace{10 mm}\forall p_{uv}^{ij} \in P
\label{bigM1}
\end{equation}
\begin{equation}
{f_{uv}^{ij}} - D_{ij} \chi_{v}^{j}\leq 0\hspace{10 mm}\forall p_{uv}^{ij} \in P
\label{bigM2}
\end{equation}
$$f_{uv}^{ij} = [0, D_{ij}] \hspace{10 mm} \forall p_{uv}^{ij} \in P$$
$$\chi_{u}^{i} = [0,1] \hspace{10 mm} \forall i \in N_v, \forall u \in N_s$$
\indent The first term in the objective \eqref{start1} is for link mapping, while the second term is for node mapping. Each of these terms are divided by the respective capacities to ensure that the substrate resources with more free resources are preferred. Constraint \eqref{vir1} ensures that each virtual node is mapped to a substrate node, while \eqref{sub1} ensures that any substrate node may be used at most once for a given mapping request. Constraints \eqref{dem} and \eqref{cap} represent the virtual link demand requirements and substrate link capacity constraints respectively. Specifically, \eqref{dem} states that the flow $f_{uv}^{ij}$ on path $p_{uv}^{ij}$ should carry the total demand of the virtual link $ij$, while \eqref{cap} states that the flow $f_{uv}^{ij}$ on path $p_{uv}^{ij}$ should be at most equal to the capacity of each substrate link on that path. From constraint \eqref{bigM1}, if $\chi_{u}^{i} == 0$ then $f_{uv}^{ij} = 0$. If $\chi_{u}^{i} == 1$ then $f_{uv}^{ij} = [0, D_{ij}]$. This is also true for \eqref{bigM2}. These constraints ensure that virtual links and virtual nodes are mapped at the same time, i.e., a flow $f_{uv}^{ij}$ $-$ using the path $p_{uv}^{ij}$ starting with meta link $iu$ and ending with meta link  $jv$ $-$ is only non-zero if the virtual node $i$ is mapped onto substrate node $u$ and $j$ is mapped onto $v$. Together, \eqref{bigM1} and \eqref{bigM2} ensure that a flow $f_{uv}^{ij}$ is only non zero if both the two end links $iu$ AND $jv$ exist.\\
\indent The formulation in \eqref{start1}$-$\eqref{bigM2} is intractable for two reasons; first, the restrictions that variables $\chi_{u}^{i}$ and $f_{uv}^{ij}$ only take on binary values, and then the fact that the number of possible paths $p_{uv}^{ij}$ (and hence the number of variables $f_{uv}^{ij}$) is very large (exponential) even for moderately sized networks. Therefore, solving the problem in its current form is impractical. There are three possibilities to solving the problem;
\begin{enumerate}
\item a relaxation to the constraints on variables $\chi_{u}^{i}$ and $f_{uv}^{ij}$ to take on continuous values,
\item restricting the number of input variables $f_{uv}^{ij}$ (by restricting the number of paths $p_{uv}^{ij}$).
\item a combination of both the first two approaches.
\end{enumerate}
For the VNE problem as formulated in \eqref{start1}$-$\eqref{bigM2}, a relaxation would require careful consideration to avoid violating the requirements that both nodes and links are mapped in one-shot (since the variables $\chi_{u}^{i}$ would no longer be able to restrict the mapping of virtual nodes to particular substrate nodes), as well splitting the flows of the virtual links across multiple links. Therefore, we take the second approach, and employ path generation, which allows for the use of only a sufficiently meaningful number of paths, and adding more paths as needed until a final solution is obtained.

\section{Path Generation}
Path generation is a method that solves mathematical programs with a large number of variables efficiently. The main idea is to solve a restricted version of the program (the restricted primal problem \cite{Goemans97}) - which contains only a subset of the variables, and then (through the use of the dual problem\cite{Goemans97}) add more variables as needed. Usually, path generation involves creating an \textit{initial solution} (\textit{restricted} set of variables) which are used in the solution for the \textit{restricted} primal problem. Then, solving pricing problems (which are determined from the dual problem), allows for adding more variables to improve the initial solution, until either a final optimal solution is found, or a stopping condition is reached.

The path generation approach taken in this paper is as follows: we start by creating an initial set of paths using a two stage node and link mapping. We then use these paths to solve a dual problem, and use the pricing problems to determine a set of paths to add to the initial solution. These paths are then used to solve a restricted primal problem to obtain a final solution. Therefore, our proposal avoids the usual iteration required in a path generation approach where the primal and dual problems are solved sequentially, many times, instead preferring only to perform a single iteration. In the next subsections, we propose a method for determining the initial set of paths, derive the pricing problems, and then describe the overall algorithm proposed in this paper.

\subsection{Initial Solution}
An initial solution (Init$-$Sol) is determined as a set of paths $P'$ in the augmented \ac{SN}, with each path $p_{uv}^{ij} \in P'$ able to support the flow $f_{uv}^{ij}$ of virtual link $l_{ij}$. Each of these paths must be able to meet the \ac{VN} mapping conditions as formulated in the primal problem. Considering the example in Fig. \ref{Diagram}, since we have two virtual links, an initial solution would have two paths, one for each virtual link. Examples of these paths could be XABEZ and XACY for virtual links XZ and XY respectively. In order to determine such a path, say for virtual link XZ, the approach in this paper is as follows: we start by performing a node mapping, which for this example, would map virtual nodes X and Z onto substrate nodes A and E respectively. This step gives us the meta links XA and EZ. In this subsection, we propose a novel node mapping solution LP$-$N for determining XA and EZ. The next step involves determining the path ABE in the \ac{SN}. This is done by using Dijkstra's algorithm, with the constraint that each link on the path should have enough capacity to support the virtual link under consideration. The complete path is determined by joining meta links XA and EZ to the respective ends of ABE.\\
\subsubsection*{LP$-$N: Node Mapping}
LP$-$N is based on mathematical programming. It is formulated in such a way that mapping of any given virtual node is relatively biased towards each substrate node by a weight. The determination and use of the weights is discussed in what follows.\\\\
\textbf{Objective:} 
It is noteworthy that, essentially, LP$-$N is aimed at achieving an initial node and link mapping. As such, many other state-of-art two-step approaches \cite{Fischer13} could be used for this purpose. However, the authors could not find a previous two-stage mapping approach that simultaneously achieves  both objectives considered in our formulation: The first objective is to keep the computation time of the initial solution as low as possible by including only the possible virtual node to substrate node combinations. Secondly, as explained later in this section, we minimize the possibility of failure at the link mapping stage, by making the node mapping \emph{aware} of the link mapping stage through the use of weights $W_{i}$ and $W_{u}$.\\\\
\textbf{Variable definition:} As before, $\chi_{u}^{i}$ is a binary variable equal to 1 when the virtual node $i$ is mapped onto substrate node $u$ and 0 otherwise.\\

\begin{equation}
\Min\sum\limits_{{i \in N_v}} \sum\limits_{{u \in \Upsilon(i)}}\frac{W_{i}}{W_{u}}\chi_{u}^{i}
\label{eqn:PaGeViNE}
\end{equation}
\ST:
\begin{equation}
\sum\limits_{u \in \Upsilon(i)}\chi_{u}^{i} = 1 \hspace{5 mm}\forall {i \in N_v}
\label{vir}
\end{equation}
\begin{equation}
\sum\limits_{i \in N_v}\chi_{u}^{i} \leq 1 \hspace{5 mm}\forall {u \in N_s}
\label{sub}
\end{equation}

$$\chi_{u}^{i} = [0,1] \hspace{10 mm} \forall i \in N_v, \forall u \in N_s$$
\begin{figure}
  {\includegraphics[width=8cm, height=4cm]{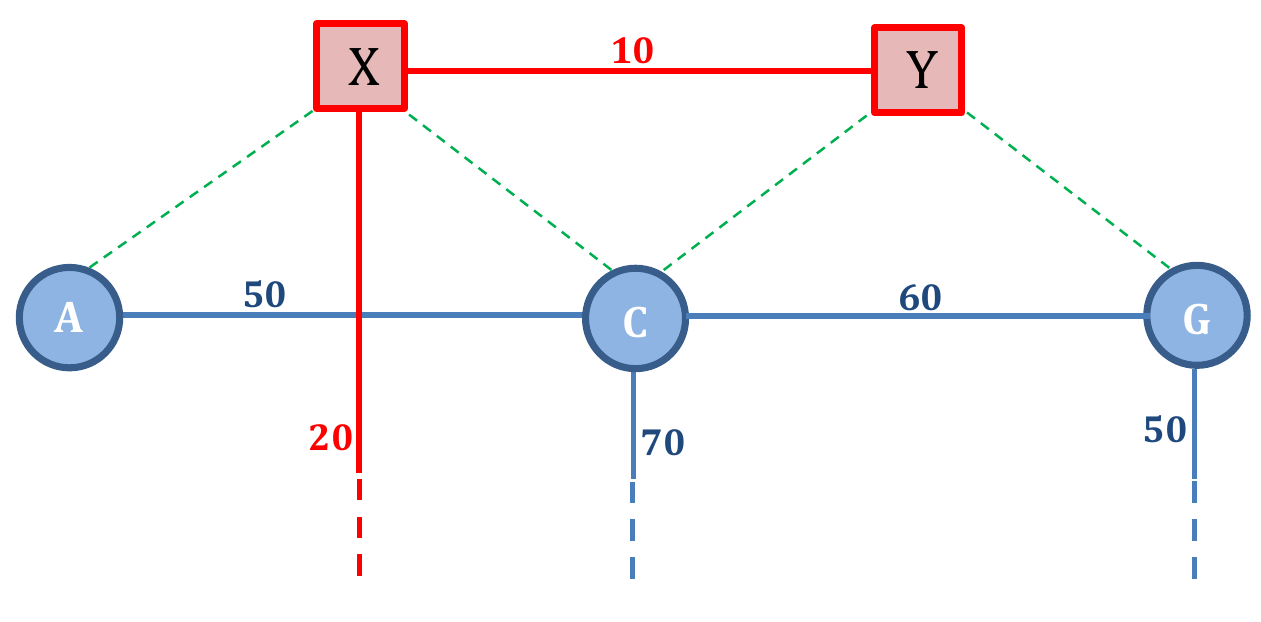}}
  \caption{Node-Link Weighted Averages}\label{Diagram1}
\end{figure}
Constraints \eqref{vir} and \eqref{sub} are the same as \eqref{vir1} and \eqref{sub1}. The weights $W_{i}$ and $W_{u}$ are dynamically determined for each virtual and substrate node respectively. The motivation to use such weights is from the need bias or coordinate the mapping of the nodes to the following link mapping step. This has been shown by related works to improve the mapping efficiency \cite{Chowdhury12} by avoiding the use of a high amount of resources for the link mapping phase. In our proposal, this is particularly important to avoid the possibility that we fail to obtain an initial solution due to unavailable \ac{SN} resources. Therefore, $W_{u}$ is defined as the weighted average of the available capacities of all the substrate links connected to $u$. Similarly, $W_{i}$ is defined as the weighted average of the demand of all the virtual links connected to $i$. To illustrate the idea behind these weighted averages, consider Fig. \ref{Diagram1}, which is a subset of the topology represented in Fig. \ref{Diagram}. The values beside each link represent the available link bandwidths and link demands respectively. As an example, considering the virtual node X,
\begingroup
\fontsize{9pt}{9pt}
$$W_{X} = 20\times\Bigg(\frac{20}{20+10}\Bigg)+10\times\Bigg(\frac{10}{20+10}\Bigg) = 16.67.$$
\endgroup
In the same way for substrate node C,
\begingroup
\fontsize{9pt}{9pt}
$$W_{C} = 50\times\Bigg(\frac{50}{50+60+70}\Bigg)+60\times\Bigg(\frac{60}{50+60+70}\Bigg)+$$$$70\times\Bigg(\frac{70}{50+60+70}\Bigg) = 61.11$$
\endgroup
The reason for using this ratio as a weight is to ensure that those substrate nodes that are connected to many substrate links with higher available resources are usually preferred, and that in case two or more virtual nodes have a given substrate node in their possible node set (such as X and Y in Fig. \ref{Diagram}), then the substrate node would always be allocated to that virtual node with the highest weighted average link demand. This achieves some level of coordination between the node mapping and link mapping phases and thereby reduces the probability of rejecting link mapping requests.\\
\indent We note that there could be instances where the weighted averages lead to selecting substrate nodes with less good links, especially when the links have widely differing residual capacities. For example, a node connected to two links with residual capacities $80$ and $10$ respectively will have a $W_1 = 72$, while a node connected to two links with residual capacities $60$ and $70$ respectively will have a $W_2 = 65$. In this case, the first node will be selected yet the second node \emph{could} be a better choice. One simple solution to handle such scenario is to use the sum of two averages: the weighted average and a simple average. However, it is worth mentioning that in our approach network embedding is done in such a way that the average loads of \ac{SN} nodes and links are balanced, this way, avoiding scenarios where some node and/or links have widely differing residual capacity. The procedure, \textbf{Init-Sol}, for determining the initial solution is shown in Algorithm \ref{Algorithm 1}.

\begin{algorithm}[t]
\caption{Init$-$Sol $(G_v(N_v,L_v), G_s(N_s,L_s))$}
\label{Algorithm 1}
\begin{algorithmic}[1]
\FOR{$i \in N_v$}
\STATE $Determine\hspace{1 mm}Candidate\hspace{1 mm}Node\hspace{1 mm}Set,\hspace{1 mm}\Upsilon(i)$
\IF{$\Upsilon{(i)} = \emptyset$}
\STATE $Reject\hspace{1 mm}Request$
\STATE $\textbf{end}$
\ENDIF
\STATE $Calculate\hspace{2 mm}W_{i}$
\ENDFOR
\FOR{$u \in N_s$}
\STATE $Calculate\hspace{2 mm}W_{u}$
\ENDFOR
\STATE $Solve:$ \textbf{LP-N}

\FOR{$l_{ij} \in L_v$}

\FOR{$u \in \Upsilon{(i)}$}

\IF{$\chi_{u}^{i} = 1$}

\STATE Meta Link 1: $l_1 = iu$
\STATE Start Node, $s = u$

\ENDIF

\ENDFOR

\FOR{$v \in \Upsilon{(j)}$}

\IF{$\chi_{v}^{j} = 1$}

\STATE Meta Link 2: $l_2 = jv$
\STATE End Node, $t = v$

\ENDIF

\ENDFOR

\STATE \textbf{LinkMapping:} $p_s = Dijkstra\Big(s, t, G_s(N_s,L_s)\Big)$
\STATE \textbf{Create Path:} $p_{uv}^{ij} = l_1+p_s+l_2$
\STATE \textbf{Add } $p_{uv}^{ij}$ \textbf{ to } $P'$

\ENDFOR
\end{algorithmic}
\end{algorithm}
\subsection{Pricing Problem}
To determine which paths should be added to the initial set so as to improve the solution, we need to solve the pricing problems for LP$-$P. In order to identify the pricing problems we first formulate the dual problem LP$-$D for the primal problem LP$-$P. The formulation of a dual problem from a primal can be obtained in five steps \cite{Lahaie08}.
\begin{enumerate}
\item Creating a dual variable for every primal constraint,
\item Creating a dual constraint for every primal variable,
\item The right-hand sides of primal constraints become coefficients for the dual objective,
\item The coefficients of the primal become right-hand sides of the dual constraints,
\item If the primal problem is a maximisation problem, the dual is a minimisation problem.
\end{enumerate}
In Table \ref{primaldual}, we summarize these steps, giving the bounds for the resulting dual constraints and variables for all possible cases of primal variables and constraints respectively. These conventions reflect the interpretation of the dual variables as shadow prices of the primal problem. A less-than-or-equal-to constraint, normally representing a scarce resource, has a positive shadow price, since the expansion of that resource generates additional profits. On the other hand, a greater-than-or-equal-to constraint usually represents an external requirement (e.g., demand for a given resource). If that requirement increases, the problem becomes more constrained; this produces a decrease in the objective function and thus the corresponding constraint has a negative shadow price. Finally, changes in the right hand side of an equality constraint might produce either negative or positive changes in the value of the objective function. This explains the unrestricted nature of the corresponding dual variable.\\\\
\textbf{Dual Variables definitions:}
To determine the dual program, we start by relaxing the bounds of the variables $\chi_u^i$ and $f_{uv}^{ij}$ such that $\chi_u^i \geq 0$ and $f_{uv}^{ij} \geq 0$. Then, following the five steps stated above, we define six dual variables as follows: $\lambda_i$  for the virtual node constraints \eqref{vir1}, $\mu_{ij}$ for the virtual links demand constraints in \eqref{dem}, $\eta_u >= 0$ substrate node constraints in \eqref{sub1}, $\gamma_{uv} >= 0$ substrate links available capacity constraints in \eqref{cap}, $\sigma_{iu} >= 0$ for simultaneous node and link mapping constraint constraint \eqref{bigM1} and $\tau_{jv} >= 0$ for constraint \eqref{bigM2}. Since most results of duality for linear programs do extend to integer programming\cite{Guzelsoy10}, the dual formulation in this paper is based on \cite{Lahaie08}.

The \textbf{objective} of the dual formulation \eqref{dualstart}$-$\eqref{price2} is to obtain a mathematical program that produces a maximized value as close as possible to that of its original primal program for any instance of the variables. Therefore, the dual of the primal formulation in \eqref{start1}$-$\eqref{bigM2} is:
\begin{table}[t]
\renewcommand{\arraystretch}{1.5}
\caption{Relationship between dual and primal problems}
\label{primaldual}
\centering
\rowcolors{2}{gray!25}{white}
\begin{tabular}{l||l}
\hline
\bfseries Primal & \bfseries Dual\\
\hline\hline
\textbf{Objective Function} & \\
Maximisation & Minimisation\\ \hline
\textbf{Variable bounds} & \textbf{Constraint bounds}\\
$-\infty \leq i \geq +\infty $ &  $i =$\\
$i \geq 0$ &  $i \geq $\\
$i \leq 0$ &  $i \leq$\\ \hline
\textbf{Constraint bounds} & \textbf{Variable bounds}\\
$j  = $ &  $ -\infty \leq j \geq +\infty $\\
$j \geq $ &  $j \leq 0  $\\
$ j \leq  $ &  $j \geq 0 $\\
\hline
\end{tabular}
\end{table}

\begin{equation}
\Max \sum \limits_{i \in N_v}\lambda_i  + \sum \limits_{l_{ij} \in L_v}D_{ij}\mu_{ij} - \sum \limits_{u \in N_s}\eta_u - \sum \limits_{l_{uv} \in L_s}A_{uv}\gamma_{uv}
\label{dualstart}
\end{equation}

$\ST$
\begin{equation}
\lambda_i + \sum \limits_{p_{uv}^{ij} \in P} (\sigma_{iu} + \tau_{jv}) - \eta_u \leq \frac{1}{A_{u}}\hspace{10 mm}\forall l_{iu} \in L_x
\label{price1}
\end{equation}

\begin{equation}
\mu_{ij} - \sigma_{iu} - \sum \limits_{l_{uv} \in p_{uv}^{ij}}\gamma_{uv} -  \tau_{jv} \leq \sum \limits_{ l_{uv}, l_{iu}\in p_{uv}^{ij}} \frac{1}{A_{uv}}\hspace{10 mm}\forall {p_{uv}^{ij} \in P}
\label{price2}
\end{equation}

\begin{figure}[t]
  {\includegraphics[width=8.5cm, height=5.5cm]{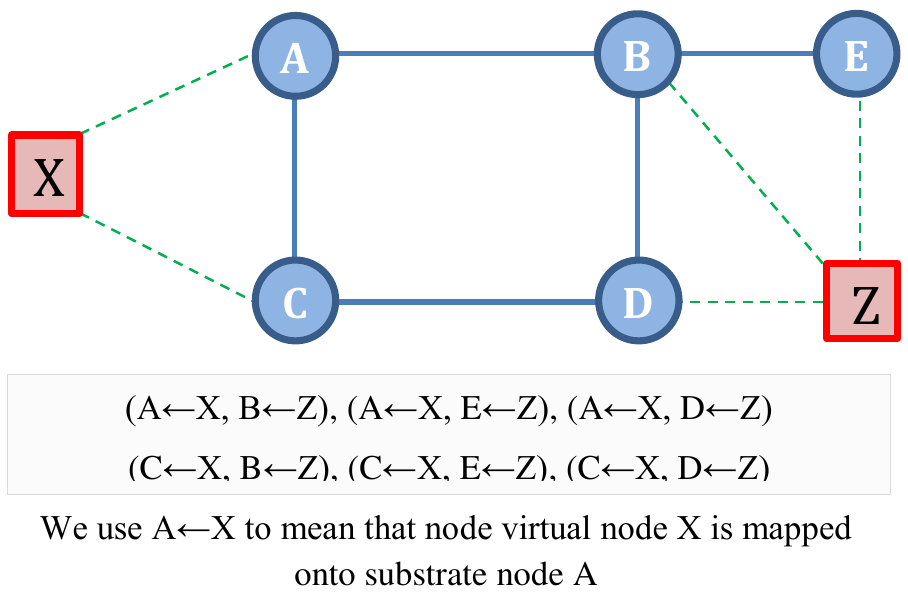}}
  \caption{Possible substrate node combinations for virtual link XZ}
  \label{Diagram2}
\end{figure}

The pricing problems are shown in \eqref{price1} and \eqref{price2}. From \eqref{price1}, the pricing condition for substrate nodes can be determined as: $$\lambda_i + \sum \limits_{p_{uv}^{ij} \in P} (\sigma_{iu} + \tau_{jv}) > \frac{1}{A_{u}} + \eta_u$$
 However, since the variables $\chi_{u}^{i}$ are much fewer compared to $f_{uv}^{ij}$, we include all the possible substrate nodes for each virtual node in the restricted primal problem. This eliminates the need for node pricing and we are left to deal with only the link pricing problem \eqref{price2}: $$\mu_{ij} > \sum \limits_{ l_{uv}, l_{iu}\in p_{uv}^{ij}} \frac{1}{A_{uv}} + (\sigma_{iu} + \sum \limits_{l_{uv} \in p_{uv}^{ij}}\gamma_{uv} +  \tau_{jv})$$
This pricing problem can be solved using the shortest path algorithm. Any path $p_{uv}^{ij} = S_{iu} + (l_{uv} \in P_{uv}) + T_{jv}$ in the augmented \ac{SN} whose length with respect to the dual variables (this means that the costs of the substrate links $l_{uv} \in P_{uv}$ are $\gamma_{uv}$, those of meta links $S_{iu}$ are $\sigma_{iu}$ and those of $T_{jv}$ are $\tau_{jv}$) is smaller than $\mu_{ij}$ satisfies the inequality above, and \emph{has the potential} to improve the solution.
However, a change in path for any given virtual link could necessitate a change in the mapping of one of its end nodes, which would change the prices and feasibility of mappings for other virtual links connected to it. For example in Fig. \ref{Diagram}, if the virtual node X is mapped onto substrate node C, all the paths for both links XZ and XY go through C. If the path for say XZ is changed to go through A, it would either mean that the path for XY should also be changed to go through A, otherwise this path cannot be used to give a feasible and improved solution. Therefore, addition of paths individually for each virtual link does not guarantee that each of the added paths would still lead to a feasible solution, and for as long as the added path cannot yield a feasible solution, this path cannot lead to improvement in the solution of the restricted primal problem. In this case, there would be no guarantee that the pricing problems can be solved in polynomial time, as it could require quite a number of iterations before enough paths are added to actually improve the solution.

In this paper, instead of adding individual paths for each virtual link in each iteration of the path generation algorithm, we include all the possible shortest path combinations after solving the formulation \eqref{dualstart} $-$ \eqref{price2}. We use Fig. \ref{Diagram2}, which is extracted from Fig. \ref{Diagram}, to illustrate this for the case of virtual link XZ. Since the node X has two possible substrate nodes and virtual node Z has three possible substrate nodes, then the possible combinations for these nodes are 6. In our pricing solution, we determine the shortest path $-$ based on the weights in \eqref{wegt2} for each of these 6 possible end node combinations.
\begin{equation}
\sum \limits_{ l_{uv}, l_{iu}\in p_{uv}^{ij}} \frac{1}{A_{uv}} + \Bigg(\sigma_{iu} + \sum \limits_{l_{uv} \in p_{uv}^{ij}}\gamma_{uv} +  \tau_{jv}\Bigg)
\label{wegt2}
\end{equation}
This is done for all the virtual links, and all the corresponding paths are added to the restricted primal problem. However, the number of paths added for each pricing iteration would be too big to handle if many iterations are carried out. Even the Dijkstra algorithm takes quite some time to find the shortest paths. For this reason, we perform only one round for the substrate paths and use the resulting shortest paths based on the dual problem to solve LP$-$P to obtain the final solution. As we show in the simulation results, the solution obtained is near optimal.

\begin{algorithm}[t]
\caption{ Final$-$Sol$(G_v(N_v,L_v), G_s(N_s,L_s))$}
\label{ColGen}
\begin{algorithmic}[1]
\STATE Create Augmented Substrate Network
\STATE Initial Paths Set: $P'\leftarrow Solve \textbf{ Init$-$Sol}$
\STATE $Solve \textbf{ LP$-$D($P'$) }$
\FOR{$l_{ij} \in L_v$}
\FOR{$u \in \Upsilon(i)$}
\FOR{$v \in \Upsilon(j)$}
\STATE $P' \leftarrow (P' + GetShortestPath(i,u,v,j))$
\ENDFOR
\ENDFOR
\ENDFOR
\STATE $Solve \textbf{ LP$-$P($P'$) }$
\end{algorithmic}
\end{algorithm}

\begin{figure*}[ht!]
    \centering
        \begin{subfigure}[t]{0.4\textwidth}
        \centering
        \includegraphics[width=7.5cm, height=4.5cm]{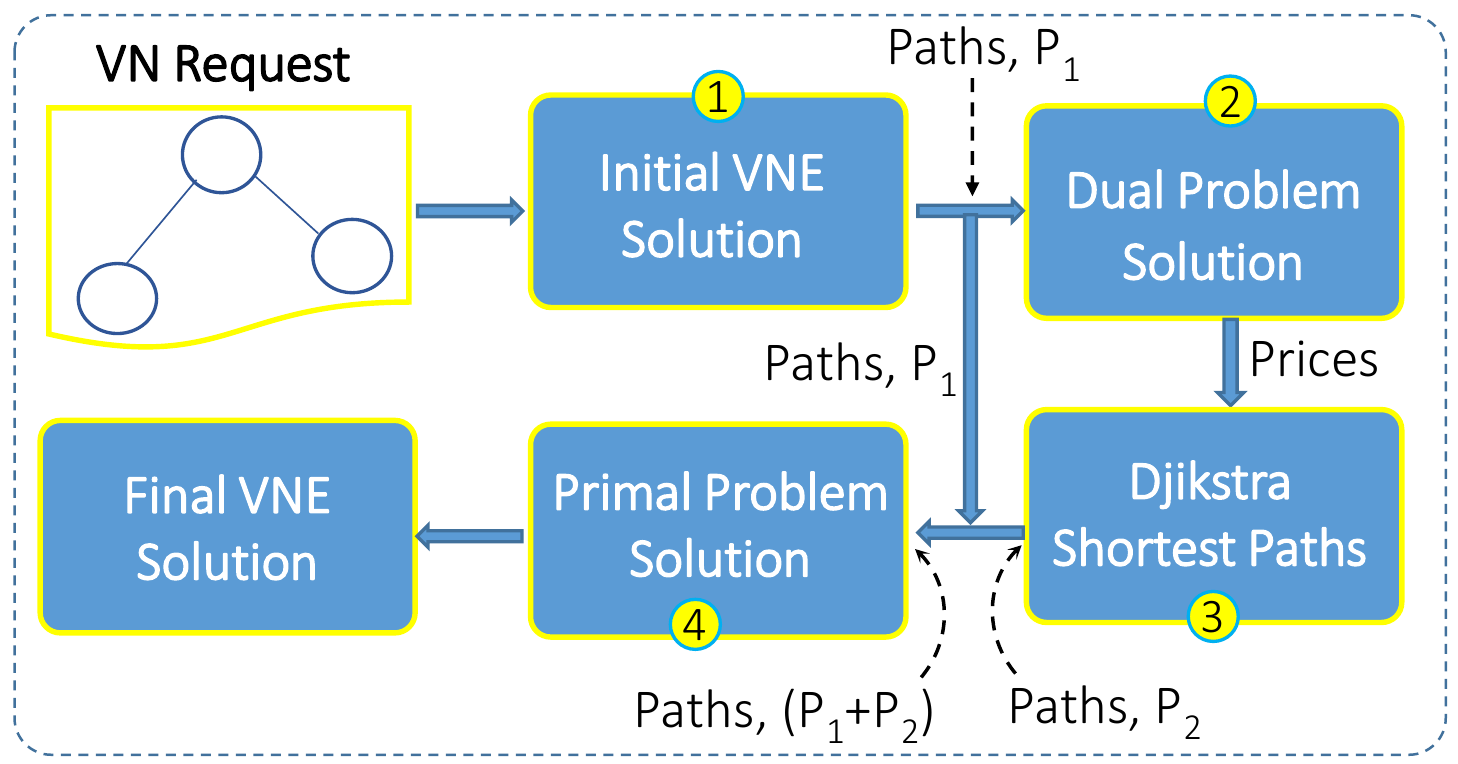}
        \caption{Summary of Path generation-based VNE Approach}
        \label{cgfig}
    \end{subfigure}
     ~
    \begin{subfigure}[t]{0.58\textwidth}
        \centering
        \includegraphics[width=9.5cm, height=4.5cm]{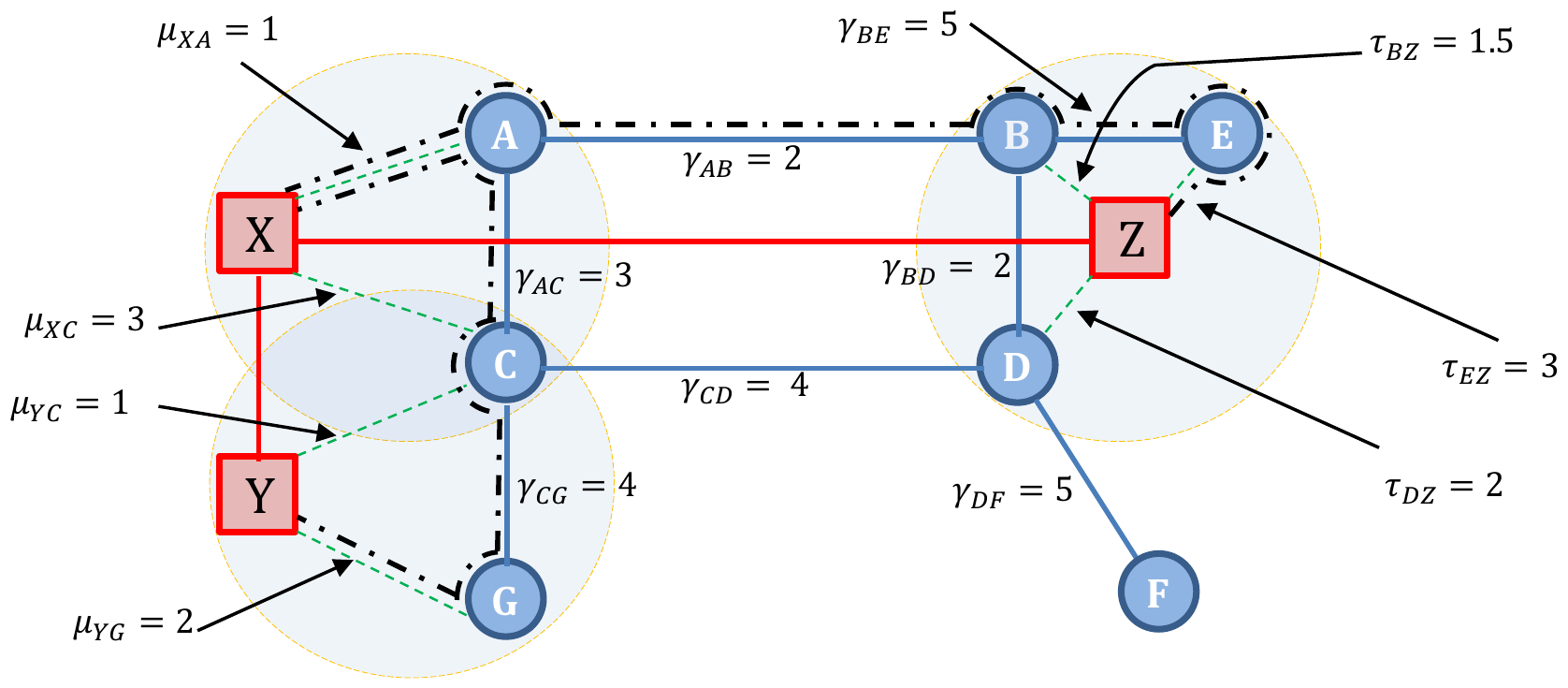}
        \caption{Initial Solution and Dual Pricing of Links}
        \label{runex2}
    \end{subfigure}   
    \caption{Running Example}
\end{figure*}

The proposed approach, \textbf{Final$-$Sol}, for determining the final solution is shown in Algorithm \ref{ColGen}.\\\\
\noindent \textbf{Example:} To illustrate the details of \textbf{Final$-$Sol} in algorithm \ref{ColGen}, we use a simple running example based on Fig. \ref{Diagram} as well as the flow diagram in Fig. \ref{cgfig}. The aim of the example is to illustrate the sequence of the proposed algorithm rather than its effectiveness, which is evaluated in the next section. As such, we keep it simple by avoiding the details of how the actual mathematical programs are solved. In Fig. \ref{runex2}, we show a possible initial solution (black dotted lines) where virtual nodes X, Y and Z have been mapped to substrate nodes A, G and E respectively. The virtual links XY and XZ have been mapped onto substrate paths ACG and ABE respectively. Therefore, based on the discussion in section IV(A), the initial solution is made up of two paths XACGY and XABEZ in the augmented substrate network. In Fig. \ref{cgfig}, these two paths make up $P_1$. With these paths, the dual problem (LP-D($P_1$)) is solved. The values of $\sigma_{iu}$, $\gamma_{uv}$ and $\tau_{jv}$ along each link in Fig. \ref{runex2} represent hypothetical values that could result from solving LP-D. As explained above, and illustrated in Fig. \ref{Diagram2}, the next step is then, for each virtual link, to find the shortest path in the substrate network for all the possible virtual-to-substrate node mappings. Using the values in Fig. \ref{runex2}, the these shortest paths are determined using Dijkstra's algorithm as shown in Table \ref{exrun}\footnote{The reader should note that for representation simplicity, the terms $1/A_{uv}$ in \eqref{wegt2} are not included in the shortest path summations in Table \ref{exrun}. These terms represent the reciprocal of the available bandwidth on each link along the shortest path in the augmented substrate network.} for the 6 combinations (see Fig. \ref{Diagram2}) of the virtual link XZ. Using a similar process, the paths corresponding to the virtual link XY are determine. These paths (excluding those which were already in the initial solution such as XABEZ) constitute $P_2$ in Fig. \ref{cgfig}.  The combined paths $(P_1 + P_2)$ are then used as inputs to solve a restricted primal problem to obtain a final solution.

\begin{table}[ht!]
\renewcommand{\arraystretch}{1.5}
\caption{Shortest Paths for Virtual Link XZ}
\label{exrun}
\centering
\rowcolors{2}{gray!25}{white}
\begin{tabular}{l c c}
\hline
\bfseries Path & \bfseries Values Along Path & \bfseries Total Path Length\\
\hline\hline
XABEZ & $1 + 2 + 5 + 3 $ & $11$\\
XABZ & $1 + 2 + 1.5 $ & $4.5$\\
XABDZ & $1 + 2 + 2 + 2 $ & $7$\\
XCZBEZ & $3 + 3 + 2 + 5 + 3 $ & $16$\\
XCABZ & $3 + 3 + 2 + 1.5 $ & $9.5$\\
XCDZ & $ 3+ 4 + 2 $ & $9$\\
\hline
\end{tabular}
\end{table}
\begin{table}[t!]
\renewcommand{\arraystretch}{1.6}
\caption{Brite Network Topology Generation Parameters}
\label{brite}
\centering
\rowcolors{2}{gray!25}{white}
\begin{tabular}{l||l||l}
\hline
\bfseries Parameter & \bfseries Substrate Network & \bfseries Virtual Network\\
\hline\hline
Name (Model) & Router Waxman & Router Waxman\\
Number of nodes (N) & $100$ and $20$ & $[15$-$25]$ and $[3$-$10]$\\
Size of main plane (HS) & $500$ & $500$\\
Size of inner plane (LS) & $500$ & $500$\\
Node Placement & Random & Random \\
GrowthType & Incremental & Incremental\\
Neighbouring Nodes & $3$ & $2$\\
alpha (Waxman Parameter) & $0.15$ & $0.15$\\
beta (Waxman Parameter) & $0.2$ & $0.2$\\
BWDist & Uniform & Uniform\\
\hline
\end{tabular}
\end{table}
\begin{table}[t!]
\renewcommand{\arraystretch}{1.5}
\caption{Performance Quality Evaluation Algorithms}
\label{table1}
\centering
\rowcolors{2}{gray!25}{white}
\begin{tabular}{l||l}
\hline
\bfseries Code & \bfseries Mapping Method\\
\hline\hline
GNMSP & Greedy Node Mapping and Shortest Path (SP) for Links\cite{Yu08}\\
CNMMCF & Coordinated Node and MCF for Link Mapping\cite{Chowdhury12}\\
VNA-1  & One-Shot Mapping\cite{Zhu06}\\
TANMSP & Topology-aware Node Mapping and SP for Links \cite{ChengXiang2011} \\
PaGeViNE & Path Generation based Virtual Network Embedding\\
ViNEOPT & Link based Optimal Virtual Network Embedding\\
\hline
\end{tabular}
\end{table}

\section{Performance Evaluation}
\subsection{Simulation Setup}
To evaluate the performance of our proposed approach, we implemented a discrete event simulator in Java, which uses the tool Brite \cite{Medina01} to generate substrate and \ac{VN} topologies. We used the tool ILOG CPLEX 12.4 \cite{CPLEX12.4} to solve the mathematical programs. Simulations were run on Windows $8$ Pro running on a 4.00GB RAM, 3.00GHz Processor Machine. Both substrate and \acp{VN} were generated on a $500\times500$ grid. The CPU and bandwidth capacities of substrate nodes and links are uniformly distributed between $50$ and $100$ units respectively. The CPU demand for \ac{VN} nodes is uniformly distributed between $2$ and $10$ units while the bandwidth demand of the links is uniformly distributed between $10$ and $20$ units. The parameters used in Brite to generate network topologies are shown in Table \ref{brite}. The parameters $\alpha$ and $\beta$ are Waxman-specific exponents, such that, $0 < \alpha \leq 1, 0 < \beta \leq 1, (\alpha, \beta) \in \mathbb{R}$. $\alpha$ represents the maximal link probability while $\beta$ is used to control the length of the edges. High values of alpha lead to graphs with higher edge densities while high values of beta lead to a higher ratio of long edges to short ones. The values used in this paper are the default values in the Brite router Waxman model used in \cite{Medina01}. Each virtual node is allowed to be located within a uniformly distributed distance between 100 and 150 units of its requested location. For embedding quality evaluations, two possible sets of network sizes have been used. One involves a \ac{SN} with $100$ nodes and \acp{VN} with number of nodes varied uniformly between $15$ and $25$, while the other has a \ac{SN} with $20$ nodes and \acp{VN} with number of nodes varied uniformly between $3$ and $10$. The need for different network sizes will be explained in a later subsection. For these simulations, we assumed Poisson arrivals at an average rate of 1 per 3 time units. The average service time of the requests is 60 time units and assumed to follow a negative exponential distribution. The experiments are performed for 1500 arrivals. For the time complexity evaluation, the number of nodes for the \ac{SN} is gradually increased from 20 to 100, and each simulation setup is repeated 20 times and average values determined.

\begin{figure*}[t]
\setlength{\abovecaptionskip}{7pt plus 0pt minus 0pt}
\setlength{\belowcaptionskip}{7pt plus 0pt minus 0pt}
\begin{minipage}{.33\textwidth}
\centering
\resizebox{.99\textwidth}{!}
{\includegraphics{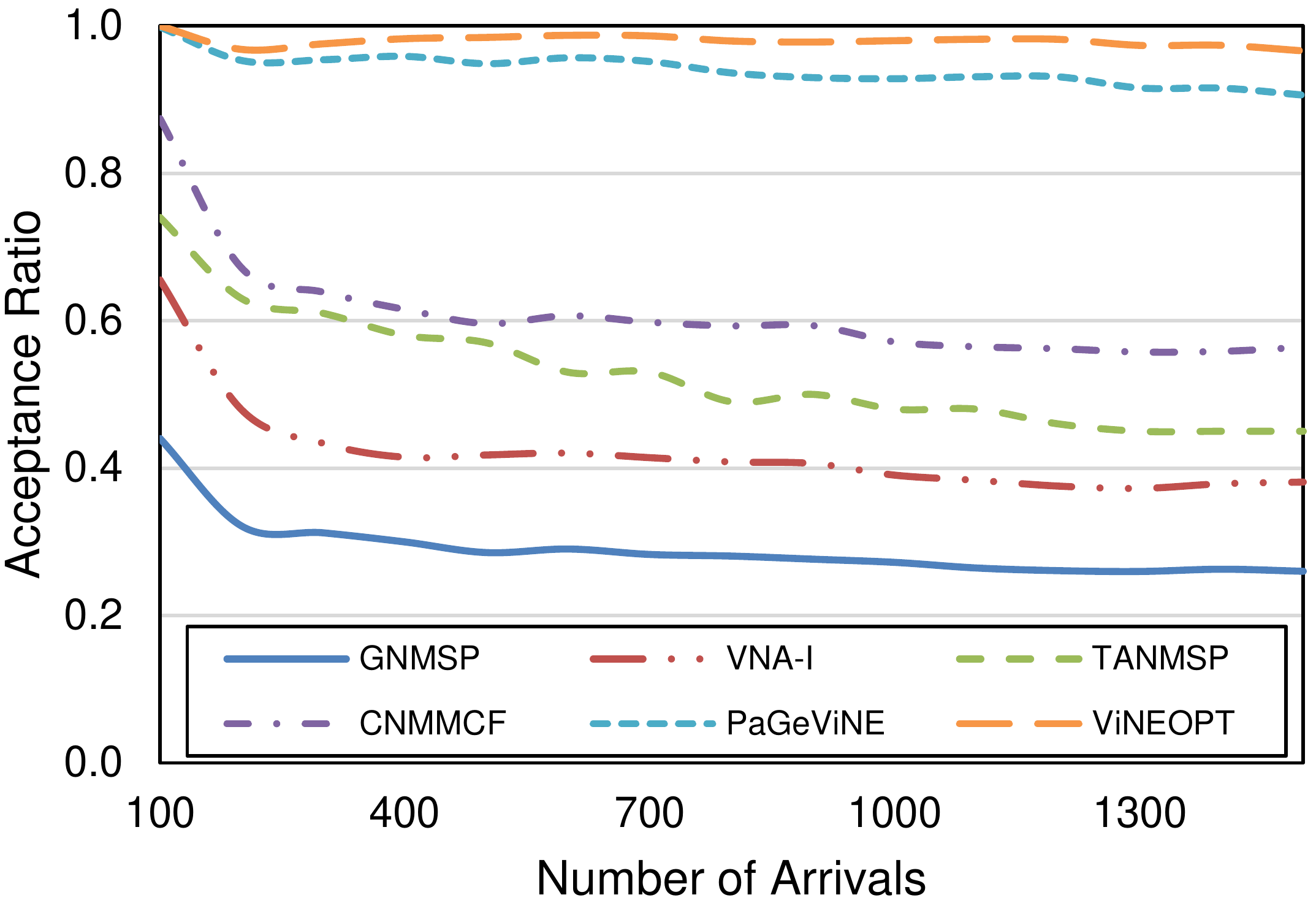}}
  \caption{Average Acceptance Ratio - 20 SN Nodes}\label{AcceptanceRatio1}
\end{minipage}
\begin{minipage}{.33\textwidth}
\centering
\resizebox{.99\textwidth}{!}
{\includegraphics{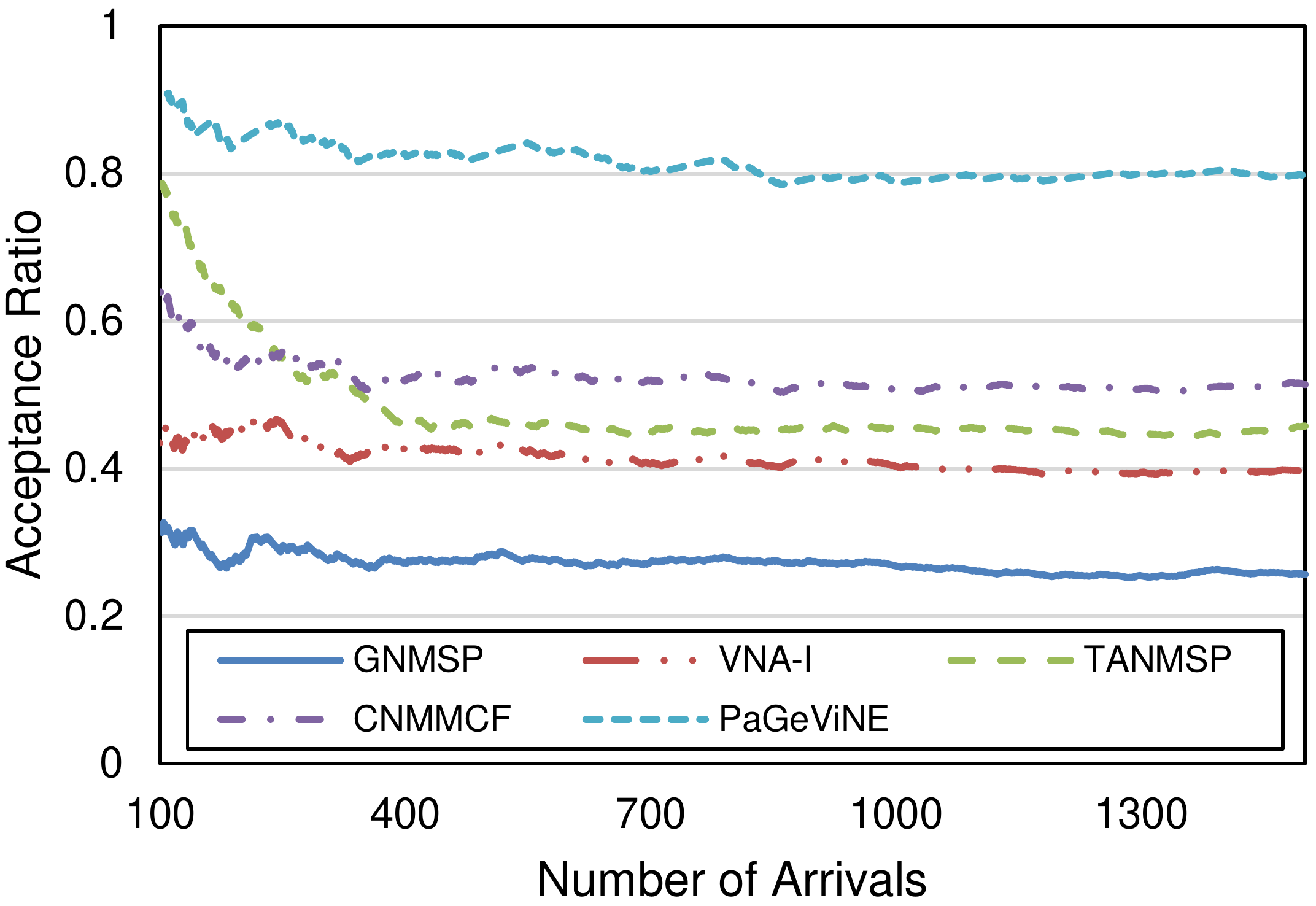}}
  \caption{Average Acceptance Ratio - 100 SN Nodes}\label{AcceptanceRatio2}
\end{minipage}
\begin{minipage}{.33\textwidth}
\centering
\resizebox{0.99\textwidth}{!}
{\includegraphics{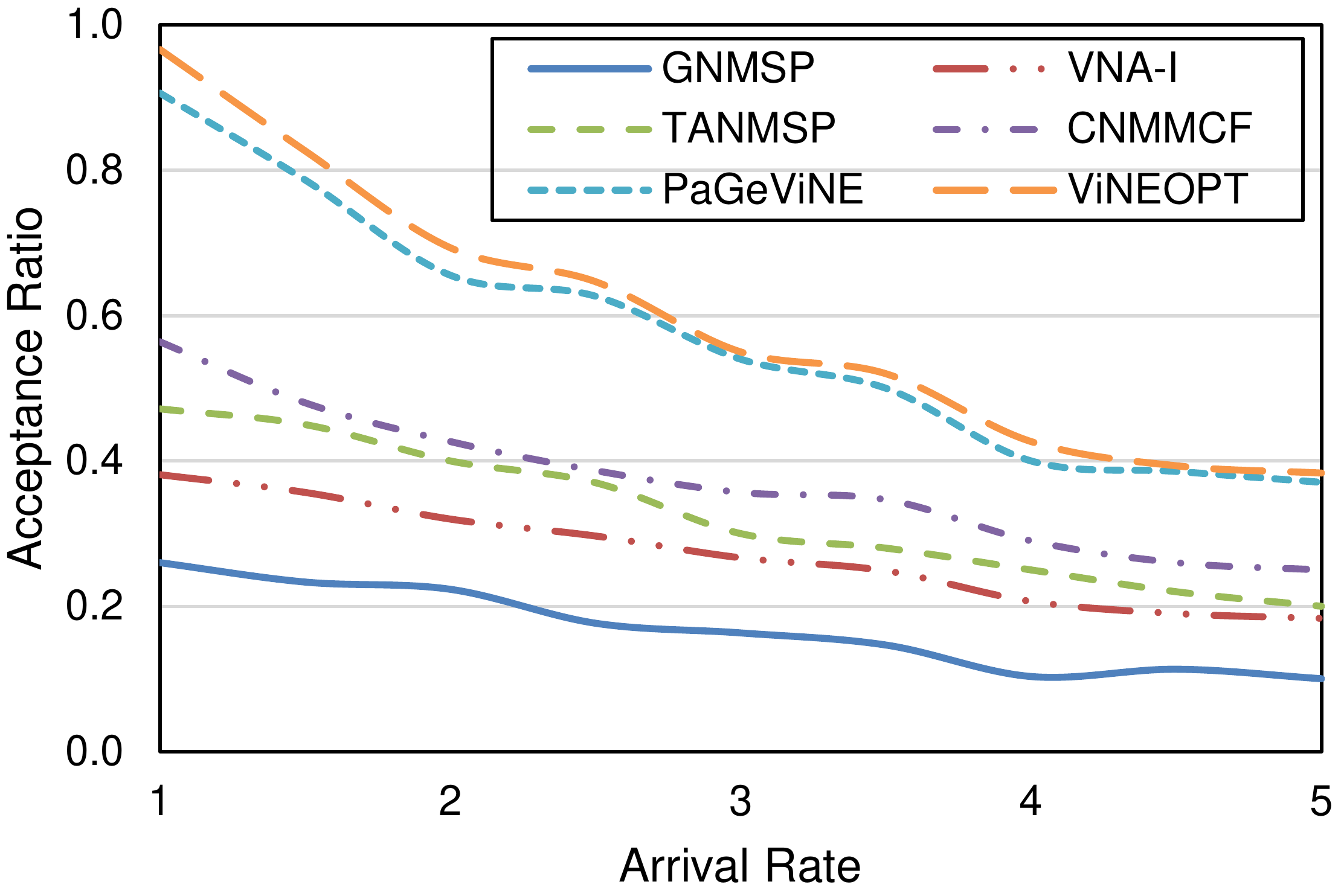}}
  \caption{Effect of VN Arrival Rate on Acceptance Ratio}
  \label{arrivalrate}
\end{minipage}
\begin{minipage}{.33\textwidth}
\centering
\resizebox{0.99\textwidth}{!}
{\includegraphics{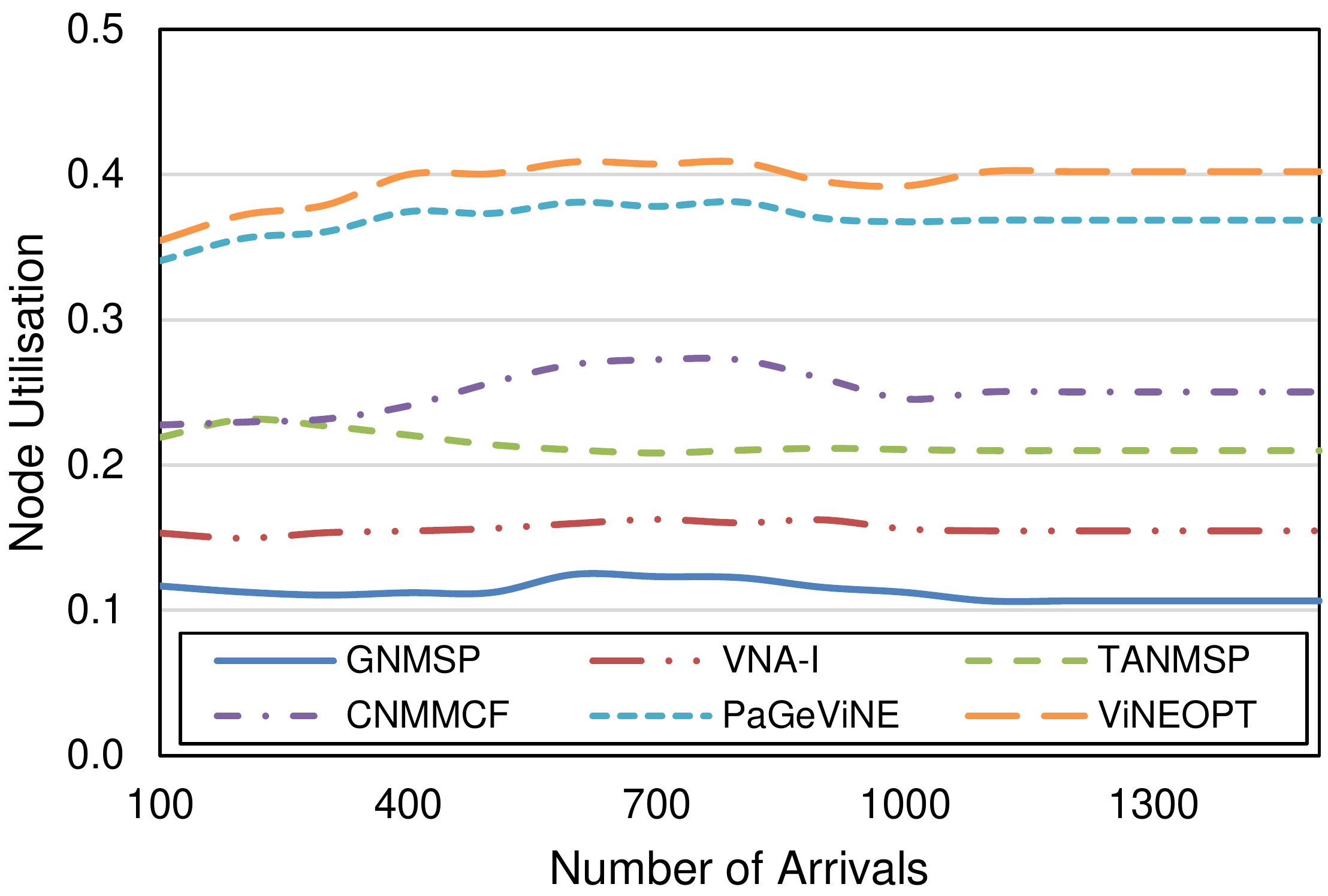}}
  \caption{Average Node Utilisation}
  \label{NodeUtilization}
\end{minipage}
\begin{minipage}{.33\textwidth}
\resizebox{.99\textwidth}{!}
{\includegraphics{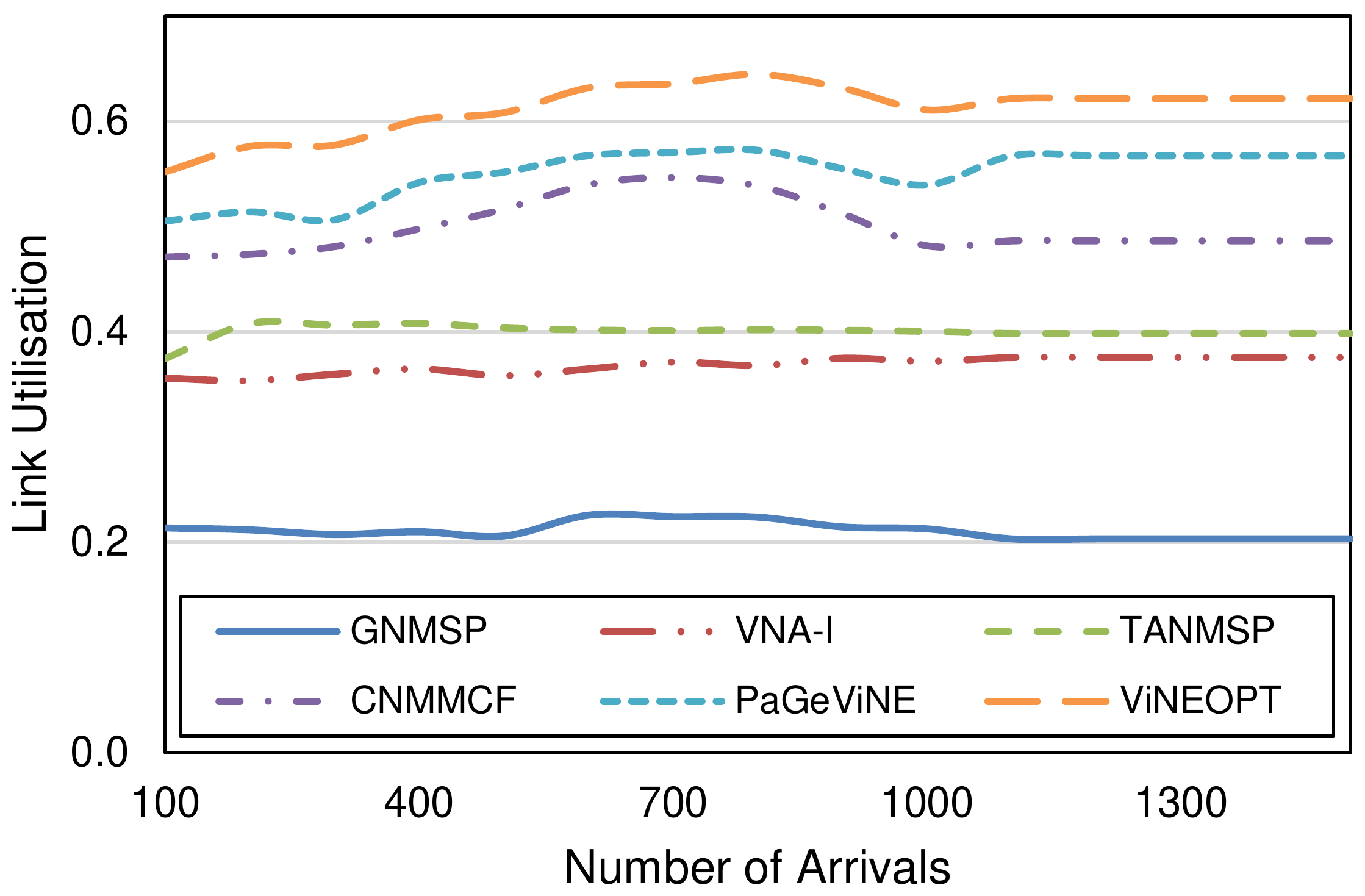}}
  \caption{Average Link Utilisation}
  \label{LinkUtilization}
\end{minipage}
\begin{minipage}{.33\textwidth}
\resizebox{.99\textwidth}{!}
{\includegraphics{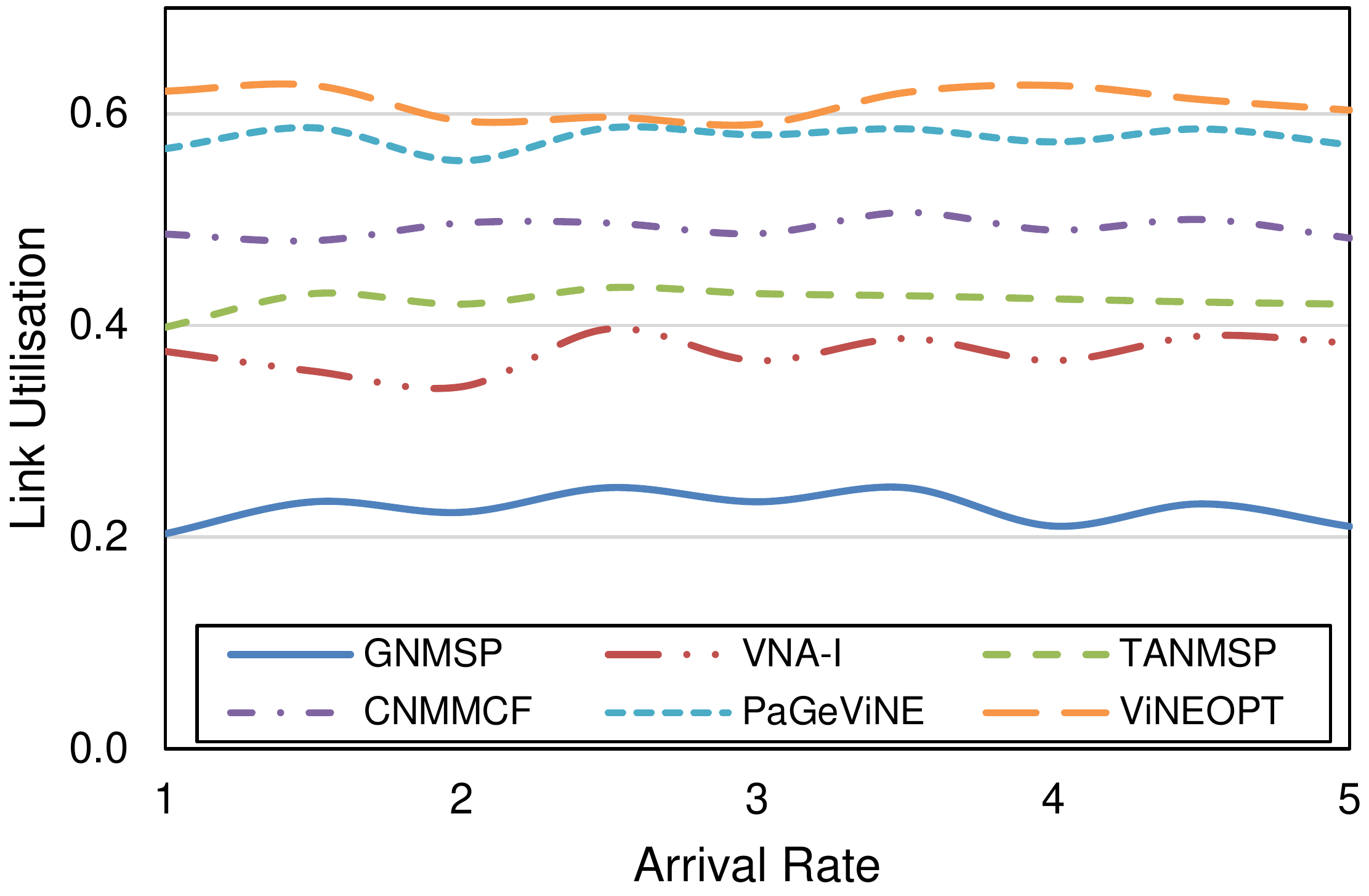}}
  \caption{Effect of VN Arrival Rate on Link Utilization}
  \label{linkarriv}
\end{minipage}

\end{figure*}

\subsection{Performance Metrics}
\subsubsection{Solution Quality} Three performance indicators $-$ Acceptance ratio, Node utilization and Link utilization $-$ are used for quality evaluation. The acceptance ratio gives a measure of the number of \ac{VN} requests accepted compared to the total requests. We define the average node utilization as the average proportion of the total substrate node capacity that is under use at any given time. In the same way, we define average link utilization as the average proportion of the total substrate link capacity that is under use at any given time.
\subsubsection{Solution Complexity} We define the time complexity of a given solution as the average time to complete the computation.
\subsubsection{Embedding Cost and Revenue}
We define the costs and revenue from embedding a given \ac{VN} the same way as a related work \cite{Chowdhury12}. In particular, we define revenue, $R\bigg(G_v(N_v,L_v)\bigg)$ as the benefit to the \ac{SN} for accepting the VN request $G_v(N_v,L_v)$. As formulated in \eqref{rev}, it is the weighted sum of the link and node demands for the \ac{VN}.
 \begin{equation}
 R\bigg(G_v(N_v,L_v)\bigg) = \sum \limits_{i \in N_v} D_{i} + \sum \limits_{l_{uv} \in L_v} D_{ij}
 \label{rev}
 \end{equation}

Similarly, in \eqref{cst}, we define an embedding cost $C\bigg(G_v(N_v,L_v)\bigg)$ as the sum of total substrate resources that are allocated to the \ac{VN} $G_v(N_v,L_v)$. $\kappa_{u}$ and $\xi_{uv}$ are parameters that represent the relative unit costs of substrate nodes and links respectively, where the virtual nodes and links are mapped.
 \begin{equation}
 C\bigg(G_v(N_v,L_v)\bigg) = \sum \limits_{i \in N_v} \kappa_{u} D_{i} + \sum \limits_{l_{ij} \in L_v} \sum \limits_{l_{uv} \in L_s} \xi_{uv} f_{uv}^{ij}
 \label{cst}
 \end{equation}
 
\subsection{Comparisons}

We compare the performance of our solution with closely related solutions. In particular, four representative solutions from the literature are chosen. We name and describe the compared solutions in table \ref{table1}. These solutions were slightly modified to fit into our formulation of the problem. Specifically, unsplittable flows, constraints on \ac{SN} capacities and constraints on virtual node locations were applied. We also implemented a baseline formulation of the optimal one-shot mapping (see Appendix). 

Since ViNEOPT requires a very long time (in excess of 1 hour for a single embedding involving a \ac{SN} of $60$ nodes and a \ac{VN} of $10$ nodes) to perform an embedding, simulations evaluating this algorithm have been restricted to \ac{SN}s with $20$ nodes and \acp{VN} with nodes from $3-10$. However, an extra simulation for acceptance ratio using larger sized networks has been performed so as to reflect more practical network sizes. This simulation excludes ViNEOPT.

\begin{figure*}[th]
\setlength{\abovecaptionskip}{7pt plus 0pt minus 0pt}
\setlength{\belowcaptionskip}{7pt plus 0pt minus 0pt}
\begin{minipage}{.33\textwidth}
\centering
\resizebox{.99\textwidth}{!}
 {\includegraphics{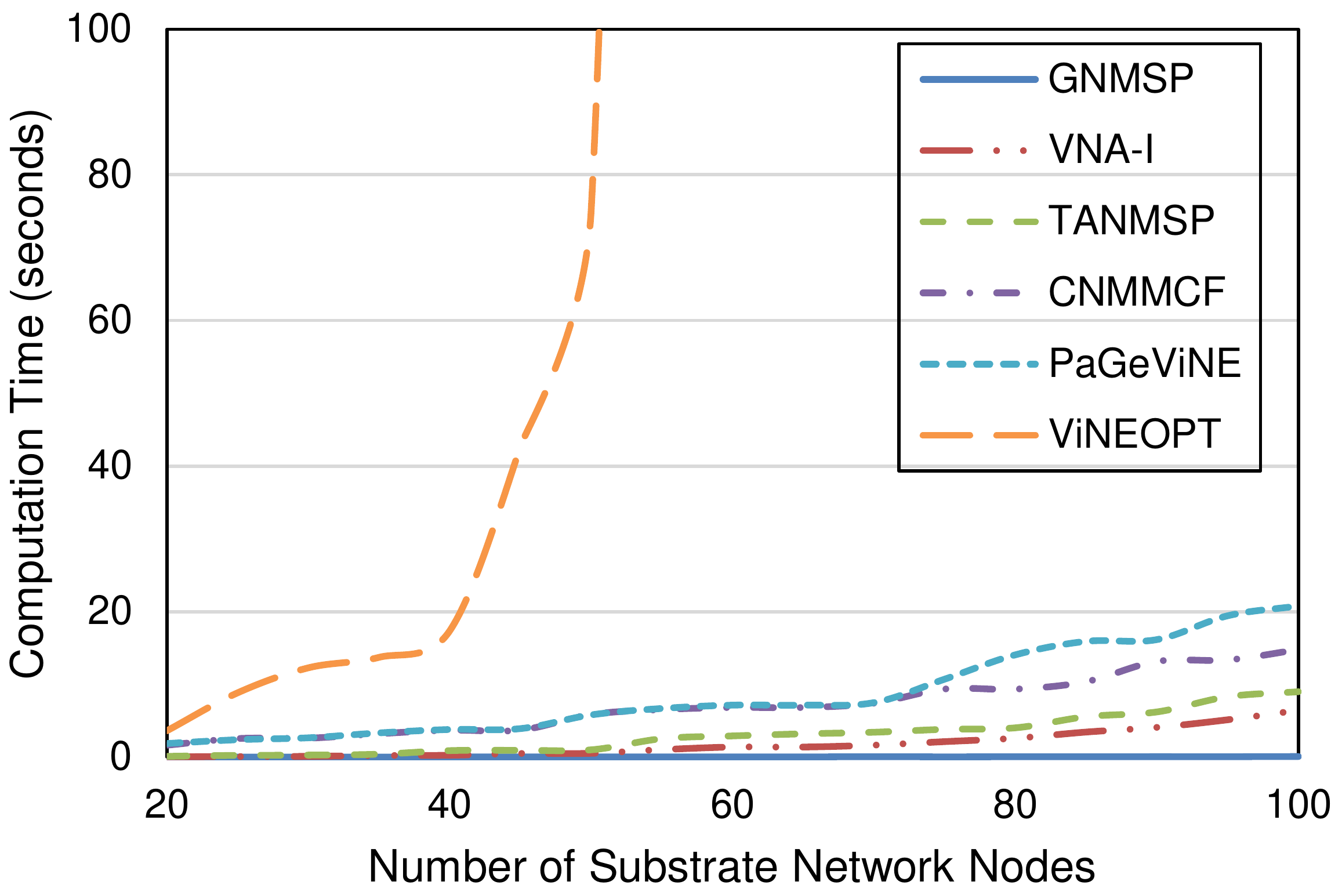}}
  \caption{Average Computation Time}
  \label{ComputationTime}
\end{minipage}
\begin{minipage}{.33\textwidth}
\resizebox{.99\textwidth}{!}
{\includegraphics{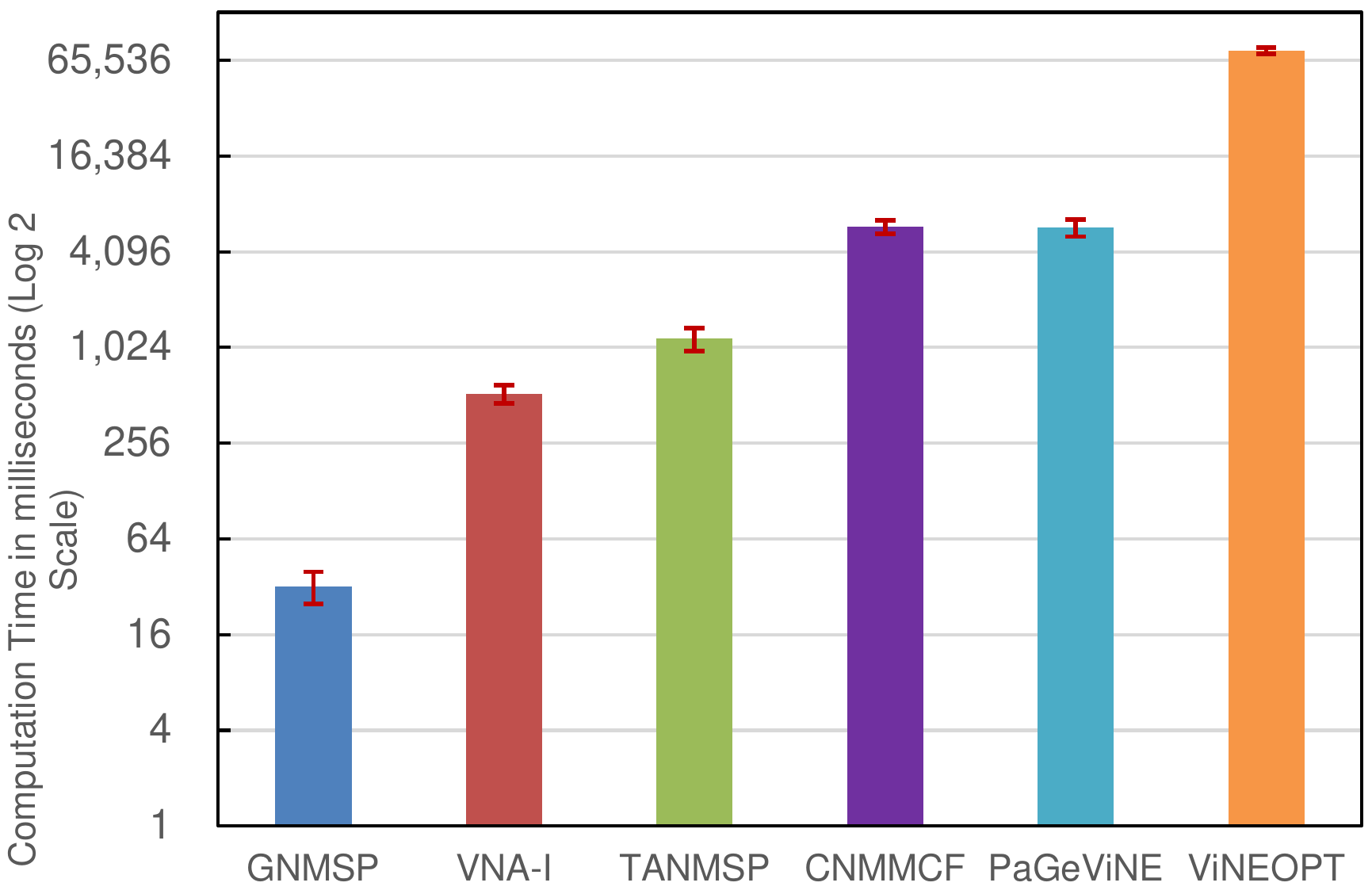}}
  \caption{95\% Confidence Interval Error Bars}
  \label{confidence}
\end{minipage}
\begin{minipage}{.33\textwidth}
\centering
\resizebox{.99\textwidth}{!}
{\includegraphics{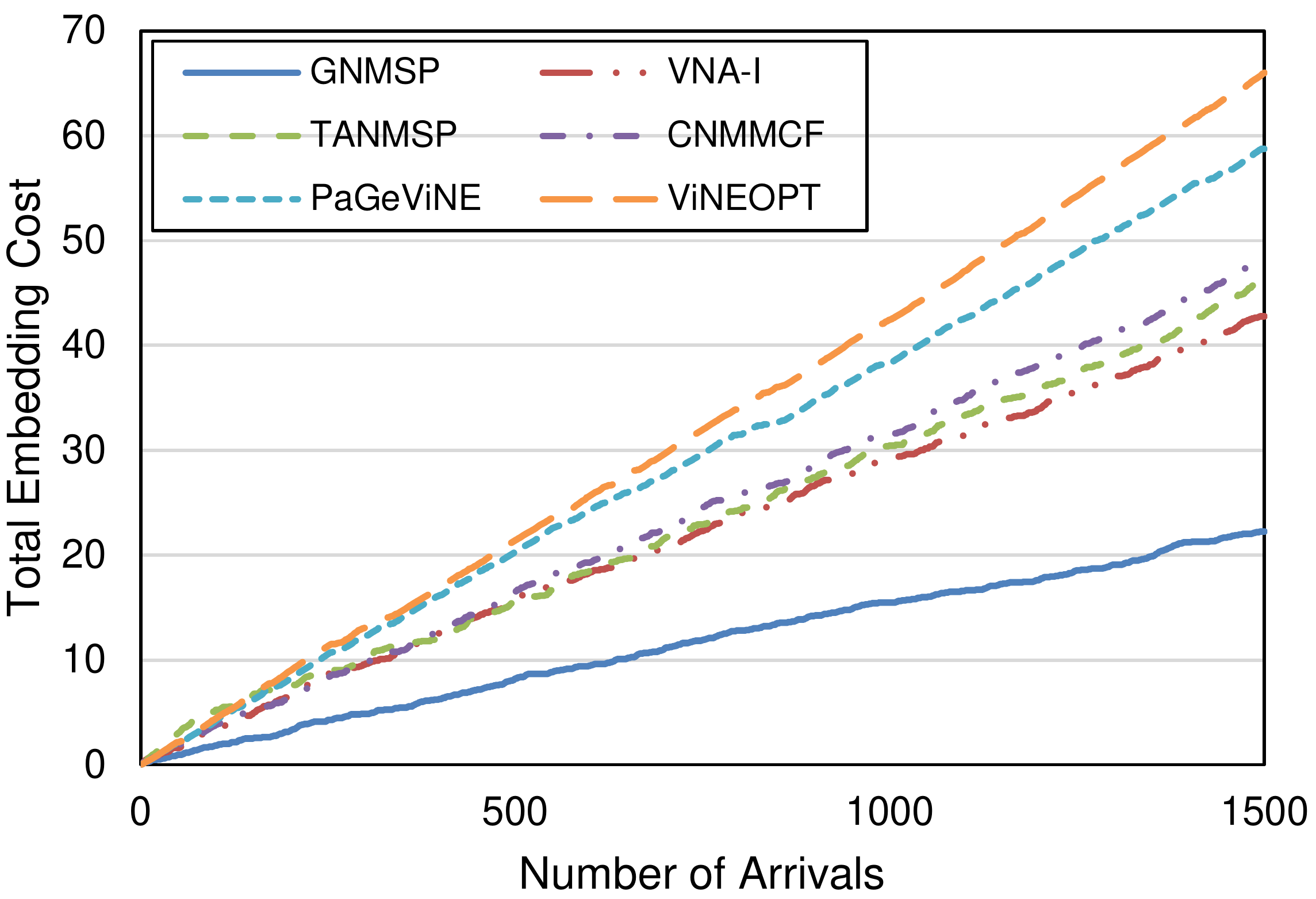}}
  \caption{Cummulative Embedding Cost}
  \label{cost}
\end{minipage}
\begin{minipage}{.33\textwidth}
\centering
\resizebox{0.99\textwidth}{!}
{\includegraphics{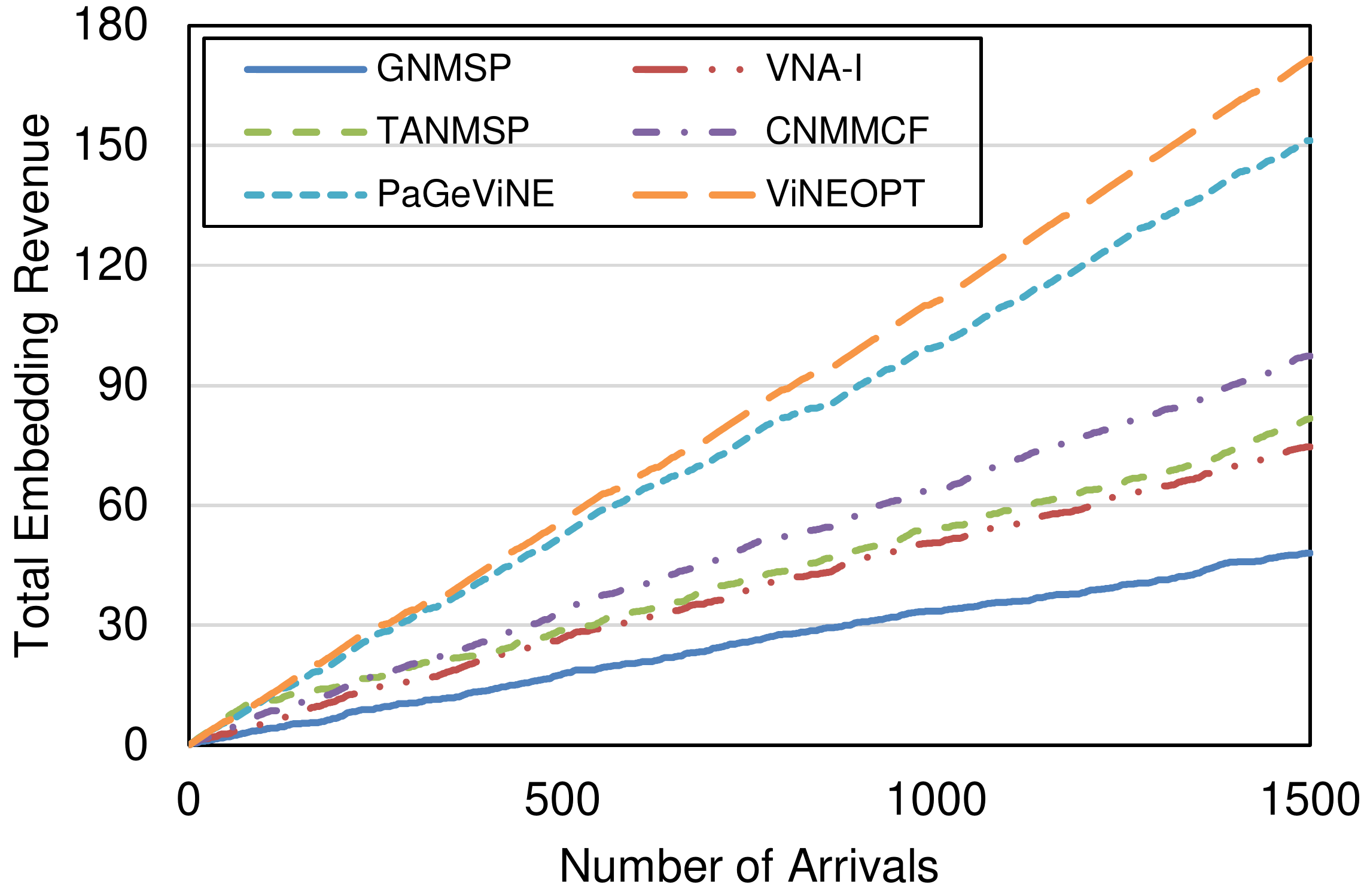}}
  \caption{Cummulative Embedding Revenue}
  \label{revenue}
\end{minipage}
\begin{minipage}{.33\textwidth}
\centering
\resizebox{0.99\textwidth}{!}
{\includegraphics{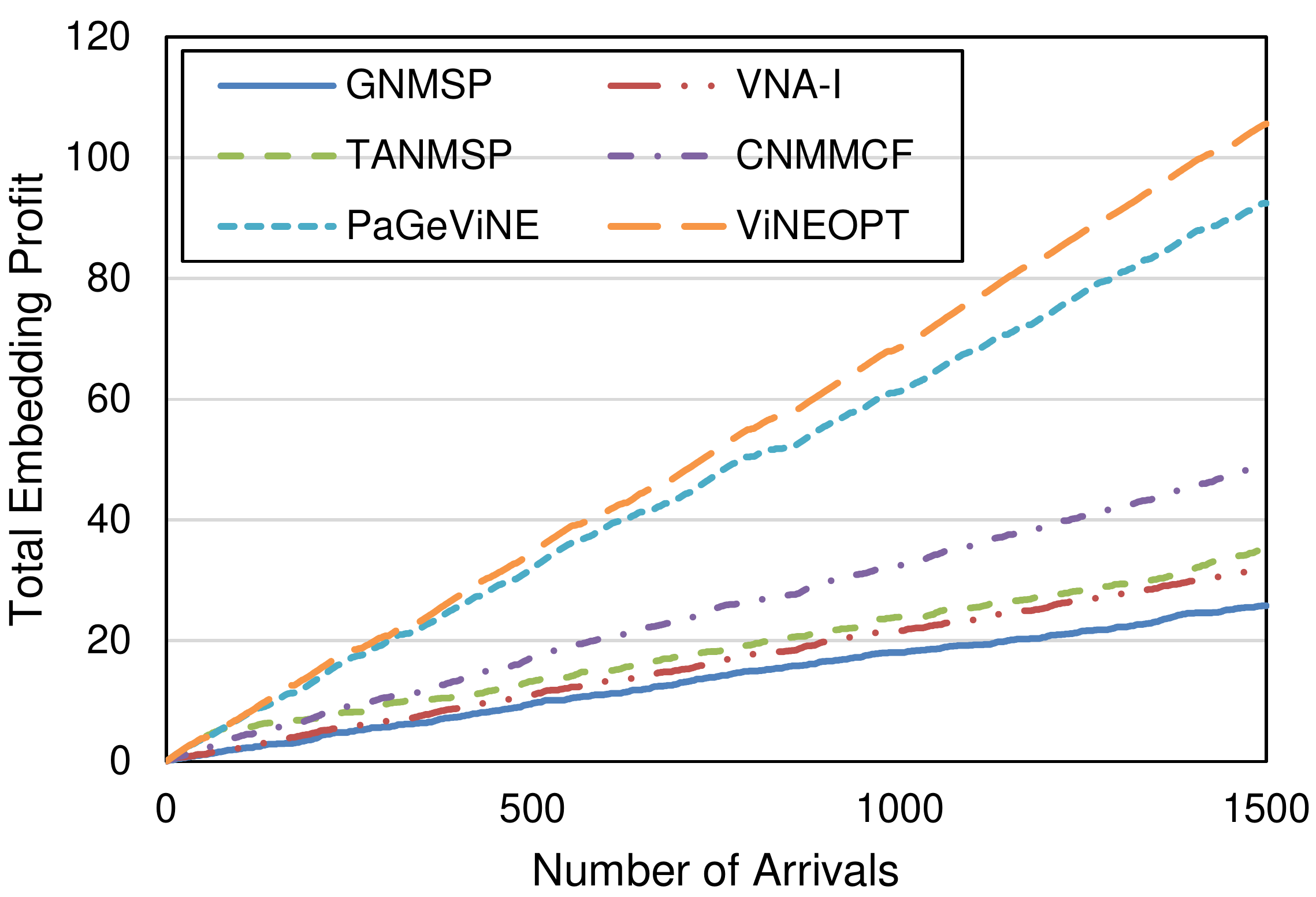}}
  \caption{Cummulative Embedding Profit}
  \label{profit}
\end{minipage}
\begin{minipage}{.33\textwidth}
\resizebox{.99\textwidth}{!}
{\includegraphics{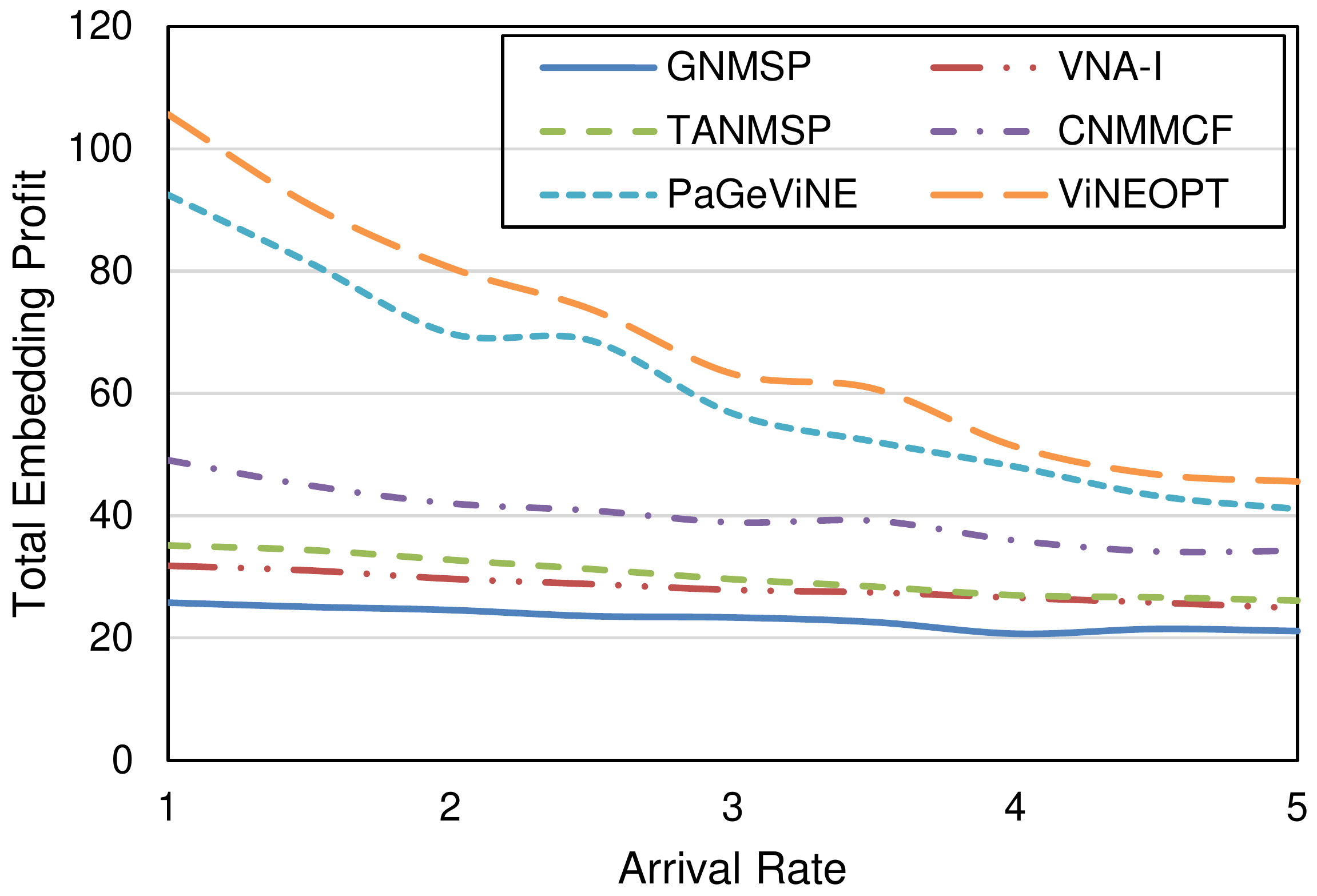}}
  \caption{Effect Arrival Rate on Profit}
  \label{profitariv}
\end{minipage}
\end{figure*}

\subsection{Results}
\subsubsection{Solution Quality}
From the graphs in Fig. \ref{AcceptanceRatio1} it is evident that PaGeViNE achieves an average acceptance ratio close to that obtained by the optimal solution ViNEOPT. In addition Fig. \ref{AcceptanceRatio1} and Fig. \ref{AcceptanceRatio2} show that PaGeViNE outperforms state-of-the-art solutions in terms of average acceptance ratio. These two figures also confirm that the embedding efficiency of PaGeViNE is not affected by increasing the size of substrate and \acp{VN}. In addition, it can be observed from Fig. \ref{arrivalrate} that even as the arrival rate of VNs increases, PaGeViNE continues to perform comparable to ViNEOPT and better than the four other approaches. The fact that CNMMCF is under-performing PaGeViNE with respect to the average acceptance ratio and resource utilization can be attributed to the fact that CNMMCF is using more resources at the link mapping stage since it performs node and link mappings separately. For VNA-1, while the node and link mapping is done in one shot, they are carried out sequentially, considering specific clusters of the \ac{SN} each time. It is therefore expected that the results would not be as good as those achieved by a global solution based on mathematical programming. It can also be observed that TANMSP, which uses the topology information to determine node mapping performs better than VNA-1 and GNMSP. However, since it still falls short of CNMMCF which determines the node mapping from a mathematical program.

It is also evident from the graphs in figures \ref{NodeUtilization} and \ref{LinkUtilization} that PaGeViNE achieves a better utilization ratio for substrate node and link resources compared to other solutions. However, we note that CNMMCF has a link utilization ratio that is comparatively close to that of PaGeViNE. Finally, it is evident from Fig. \ref{linkarriv} that the utilisation of the resources is almost un affected by the arrival rate. This confirms the fact that the rejection of VN requests is not caused by depletion of resources but rather by inefficient embedding which either fails due to bottleneck nodes. This is why mathematical programming-based algorithms which have global knowledge of the embedding perform better.

\subsubsection{Solution Complexity}
With respect to time complexity, the graphs in Fig. \ref{ComputationTime} show that the running times of GNMSP, VNA-1 and TANMSP are comparatively lower than those of PaGeViNE. Once again, this can be explained by the fact that these two solutions do not solve a mathematical program as PaGeViNE does. We also note that the computation time of PaGeViNE is slightly higher than that of CNMMCF. This can be attributed to the fact that PaGeViNE solves three mathematical programs, while CNMMCF solves only two. Moreover, it is expected that solving the problem in one-shot requires more computation than solving it in two stages, since some of the mathematical programs solved in PaGeViNE are binary. With regard to ViNEOPT we see that the computation time quickly grows exponentially. In fact, ViNEOPT could not find a solution even after 1 hour for 60 substrate nodes\footnote{Once again, this is why the simulations for acceptance ratio were split into one with $20$ \ac{SN} nodes and another with $100$ substrate nodes.}. We therefore note a significant improvement in time complexity of PaGeViNE compared to ViNEOPT. These simulations were each repeated 20 times, and the average time determined in each case. In Fig. \ref{confidence}, we show the 95\% confidence intervals of the computation time for a \ac{SN} with 50 nodes. The small error values in each of these graphs further confirms the profile in Fig. \ref{ComputationTime}. 

\subsubsection{Embedding Cost, Revenue and Profit}
Figs. \ref{cost}, \ref{revenue} and \ref{profit} show the cumulative embedding costs, revenue and profit. The profit is the difference between the revenue and cost of embedding a given \ac{VN}. We note that PaGeViNE achieves a profitability close that of ViNEOPT, which is considerably higher than that of the compared state-of-art approaches. We also note CNMMCF achieves a higher profitability than VNA-1, TANMSP and GNMSP. It is worth noting that these profiles are similar to those obtained from the acceptance ratios of the three approaches. This means that the superiority of our approach is not based on accepting \acp{VN} with less resources requirements which would be less profitable for the physical resource providers. The fact that VNA-1, TANMSP, GNMSP and CNMCMF obtained much lower embedding costs is due to rejecting most of the \ac{VN} requests, which is further confirmed by the revenue obtained, and hence profitability. In Fig. \ref{profitariv}, we evaluated the effect of the arrival rate on profitability, noting that as the arrival rate is increased, the profitability reduces. This is not surprising since an increase in the arrival rate ensures that most of the arriving VN requests in the simulation time are not accepted due to lack of resources. This profile is consistent with that of the acceptance ratio is Fig. \ref{arrivalrate}.\\

\begin{figure*}[ht!]
\setlength{\abovecaptionskip}{7pt plus 0pt minus 0pt}
\setlength{\belowcaptionskip}{7pt plus 0pt minus 0pt}
\begin{minipage}{.5\textwidth}
\centering
\resizebox{.99\textwidth}{!}
 {\includegraphics{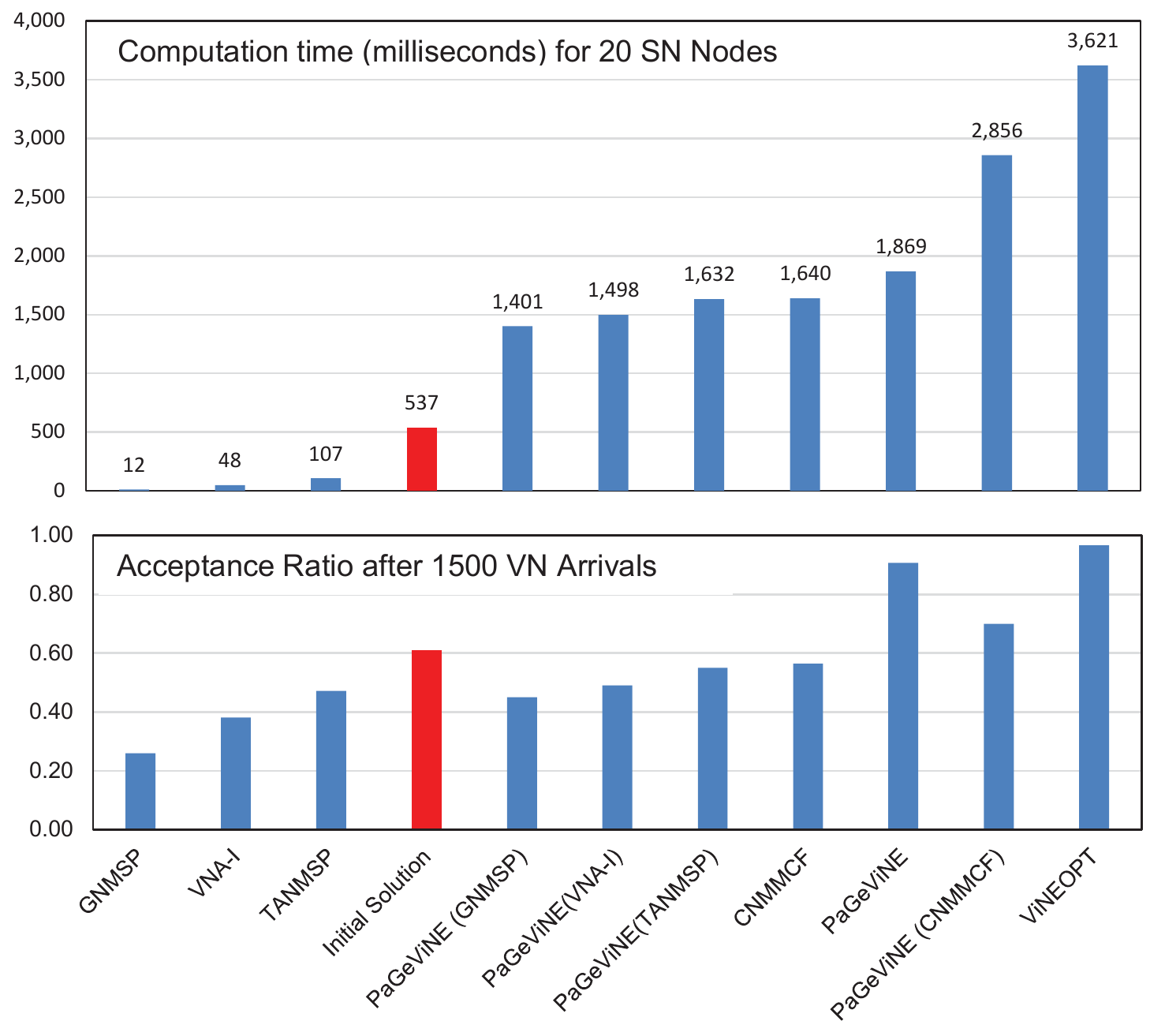}}
  \caption{Evaluation of the Initial Solution}
  \label{initsol}
\end{minipage}
\begin{minipage}{.5\textwidth}
\resizebox{.99\textwidth}{!}
{\includegraphics{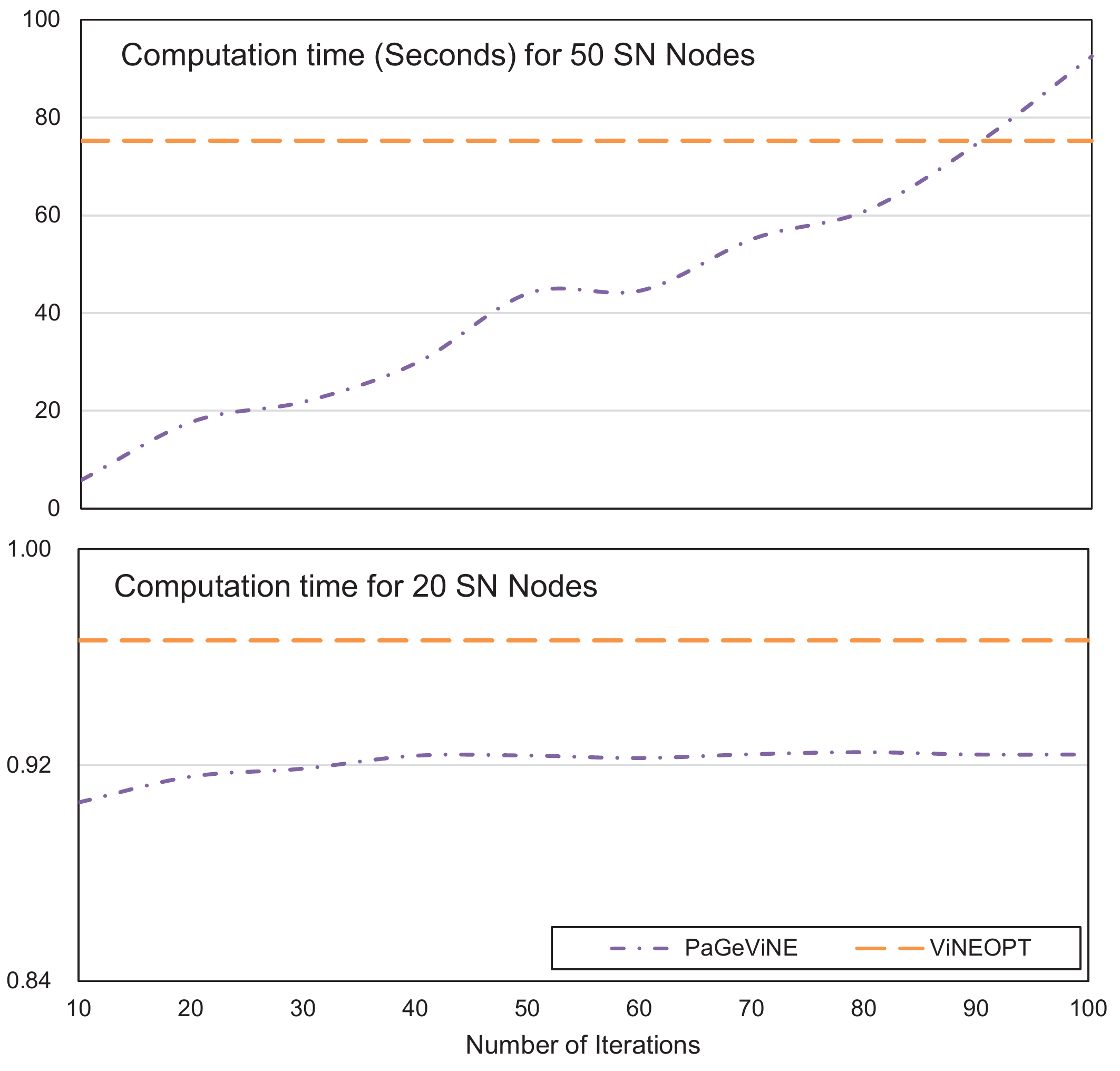}}
  \caption{Effect of Number of Iterations}
  \label{numiter}
\end{minipage}
\end{figure*}

\noindent \textbf{Effect of Initial Solution}: In Fig. \ref{initsol}, we evaluate the proposed initial solution. In particular, the effect of the initial solution on the computation time for a substrate network of 20 nodes, and the acceptance ratio after the arrival of 1,500 VN requests are shown. It can be observed that the proposed initial solution achieves the balance between time complexity and embedding quality. While it performs worse than GNMSP, VNA-1 and TANMSP in terms of computation time, it outperforms them on solution quality. Even more, it outperforms CNMMCF both on computation time as well as solution quality. The reason for this slightly better performance can be attributed to the fact that in CNMMCF node mapping is finalized by mapping each virtual node individually, which could sometimes lead to failures in embedding especially if more than one virtual node compete for a given substrate node. We also evaluated the performance of PaGeViNE in case the initial solution is changed. For example, PaGeViNE(GNMSP) means that GNMSP is used to determine the initial solution before applying path generation. These results show that PaGeViNE is dependent on an initial solution for both solution quality as well as computation time. With regard to the computation time, this dependence can be explained by the fact that the initial solution as well as the main PaGeViNE mathematical programs are solved sequentially. This means that if the computation of the initial solution takes longer, the overall solution will take longer. Similarly, since we do not allow the algorithm to run to completion, the quality of the initial solution will determine that of the final solution in two ways (1) in some cases, the initial solution just fails to even find a start solution, or (2) if the obtained initial solution is not good enough, the improvement in one iteration is not as good as it could be. These aspects are all confirmed by Fig. \ref{initsol}.\\

\noindent \textbf{Effect of Number of Iterations:} Finally, Fig. \ref{numiter} is aimed at justifying our decision to perform a single iteration rather than having an iterative approach. From the Fig. we can observe that as the number of iterations is increased, the computation time increases more rapidly than does the solution quality.

\subsection{Limitations}
The mathematical formulation \eqref{start1}$-$\eqref{bigM2} involves solving a binary program. This problem is NP-hard in the general case, and only exponential algorithms are known to solve it in practice \cite{Woeginger03}. Our approach is to reduce the number of input variables to the program using path generation. While a significant improvement in computation time is achieved compared to the optimal solution, more work can be done for instance seeking a relaxation to the program which permits to solve it in polynomial time. We however note that in practice there are \emph{high performance} tools \cite{CPLEX12.4} for solving binary programs. In particular, we have noted that the initial solution also contributes significantly to the overall computation complexity, and hence a more efficient heuristic for the same purpose could possible further enhance the results obtained in this paper. In addition, it would be interesting to make a mathematical analysis on the bounds of the computation time savings achieved in this paper.

\section{Conclusion}
In this paper we have proposed a \ac{VNE} solution which differs from previous solutions by performing node and link mappings in one shot using optimization theory and path generation. Our path generation based approach first obtains an initial solution by coordinating the node and link mapping stages, and then enhances this solution by carrying out only one round of pricing for the dual variables to obtain the final solution. Through extensive simulations, we have shown that our approach has a comparative advantage over previous approaches in terms of solution quality, achieving a comparatively superior acceptance ratio as well as \ac{VNE} revenue, which directly leads to higher profitability for \ac{SN} providers. The acceptance ratio is atleast 95\% of that obtained by the optimal solution. In addition, our approach significantly reduces solution computation time compared to the optimal one (achieving a 92\% saving in computation time for \acp{SN} of 50 nodes), and that this time complexity is comparable to that of related works.\\
\indent Looking at the future, there are several possible research avenues. With regard to time complexity, it would also be interesting to propose relaxations to the mathematical programs in order to ensure polynomial time convergence. For this purpose, we are currently investigating the feasibility of using a combination of Tabu Search and path relinking to further improve the solution time. In addition, to optimize resource allocations over time, we are exploring possibilities of modeling the substrate network state as a markovian decision process \cite{RossKW95}, and by assigning state probabilities and transition rewards be able to bias the mapping of virtual resources to more appropriate resources. Finally, we intend to extend our proposed solution to a multi-domain \ac{VNE} scenario \cite{MChowdhury10} and to consider failures in the \ac{SN} \cite{MijuSurv14, ZilongYe14}.

\section*{Acknowledgement}
The authors are grateful to the editors as well as anonymous reviewers whose insightful comments and suggestions led to a significant improvement in this paper. This work has been supported in part by FLAMINGO, a Network of Excellence project (318488) supported by the European Commission under its Seventh Framework Programme and project TEC2012-38574-C02-02 from Ministerio de Economia y Competitividad.

\appendix[ViNEOPT]
This is the link based formulation of the one-shot optimal \ac{VNE} problem. We define $f_{uv}^{ij}$ as the flow of a virtual link $l_{ij} \in L_v$ on the link $l_{uv} \in (L_s \cup L_x)$. $L_x$ is the set of all meta links in the augmented \ac{SN}.
$$\Min \sum \limits_{l_{ij} \in L_v}\sum \limits_{l_{uv} \in (L_s \cup L_x)}  \frac{1}{A_{uv}}f_{uv}^{ij}\hspace{3 mm}+\hspace{1 mm}\sum \limits_{n_v \in N_v}\sum \limits_{n_s \in N_s} \frac{1}{A_{n_s}}\chi_{n_s}^{n_v}$$
$\ST$\\\\
\textbf{Node Mapping Constraints}
$$\sum \limits_{n_s \in N_s} \chi_{n_s}^{n_v} = 1 \hspace{10 mm}\forall n_v \in N_v$$
$$\sum \limits_{n_v \in N_v} \chi_{n_s}^{n_v} \leq 1 \hspace{10 mm}\forall n_s \in N_s$$
$${f_{uv}^{ij}} - D_{ij} \chi_{u}^{i} \leq 0\hspace{10 mm}\forall uv \in L_x,\forall l_{ij} \in L_v$$
$${f_{uv}^{ij}} - D_{ij} \chi_{v}^{j} \leq 0\hspace{10 mm}\forall uv \in L_x,\forall l_{ij} \in L_v$$\\
\textbf{Capacity Constraints}
$$\sum \limits_{ij \in L_v} f_{uv}^{ij}\leq A_{uv}\hspace{10 mm}\forall l_{uv} \in (L_s \cup L_x)$$
$$\sum \limits_{uv \in L_v} f_{uv}^{ij} = D_{ij}\hspace{10 mm}\forall l_{ij} \in L_v$$\\
\textbf{Flow Conservation Constraints}\\\\
Source Nodes
$$\sum \limits_{k \in N_s} f_{ik}^{ij} - \sum \limits_{k \in N_s} f_{ki}^{ij} = D_{ij}\hspace{10 mm}\forall l_{ij} \in L_v$$
Sink Nodes
$$\sum \limits_{k \in N_s} f_{jk}^{ij} - \sum \limits_{k \in N_s} f_{kj}^{ij} = - D_{ij}\hspace{10 mm}\forall l_{ij} \in L_v$$
Intermediate Nodes
$$\sum \limits_{u \in N_s} f_{uv}^{ij} - \sum \limits_{u \in N_s} f_{uv}^{ij} =  0\hspace{10 mm}\forall l_{ij} \in L_v, \forall v \in N_s$$
Domain Constraints
$$f_{uv}^{ij} = [0, D_{ij}] \hspace{10 mm} \forall l_{ij} \in L_v, \forall l_{uv} \in (L_s \cup L_x)$$
$$\chi_{u}^{i} = [0,1] \hspace{10 mm} \forall i \in N_v, \forall u \in N_s$$

\bibliographystyle{IEEEtran}
\bibliography{IEEEabrv,biblio}

\newpage
\begin{IEEEbiography}
[{\includegraphics[width=1in,height=1.25in,clip,keepaspectratio]{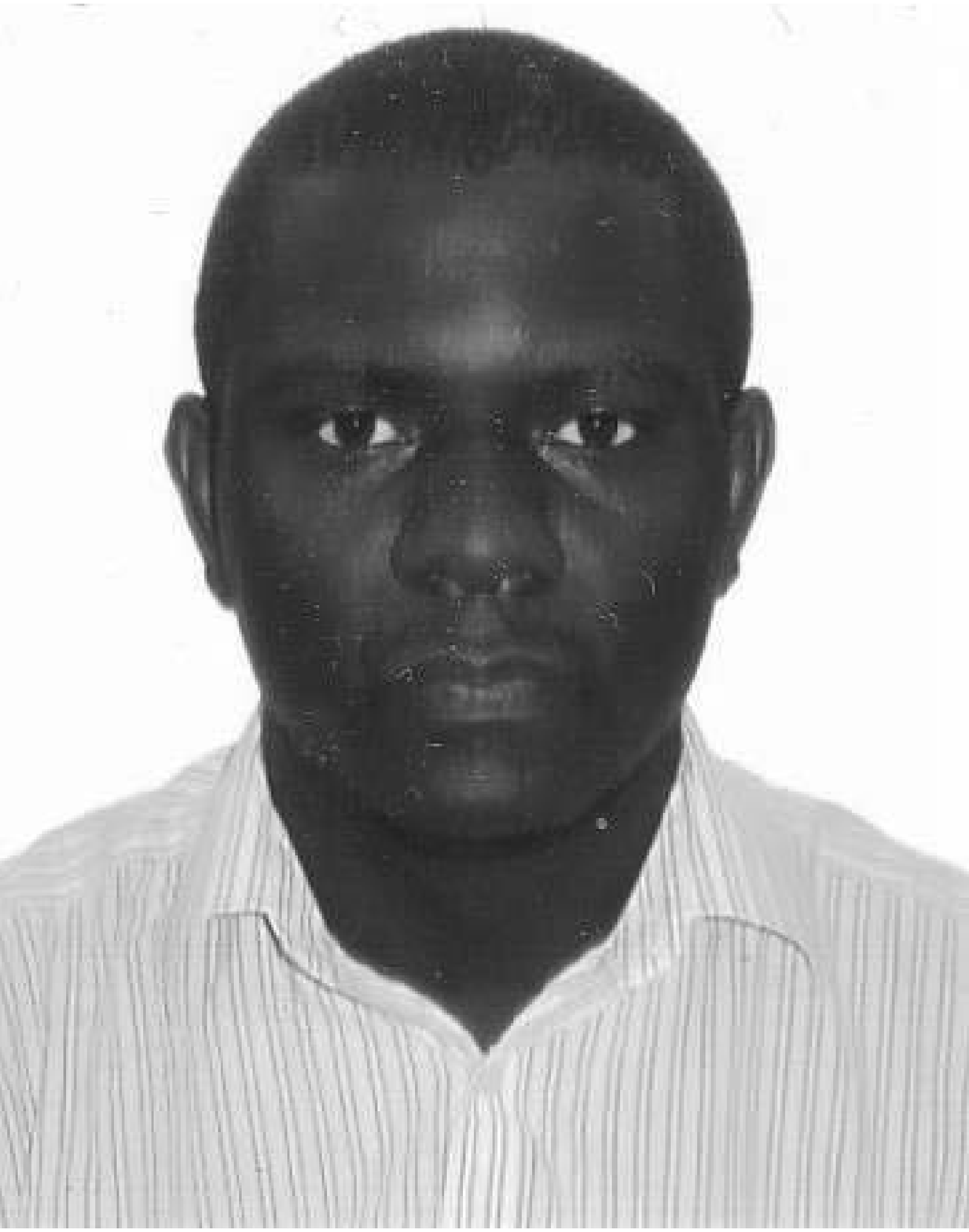}}]{Rashid Mijumbi}
obtained a Bachelors of Science Degree in Electrical Engineering from Makerere University (Kampala, Uganda) in 2009, and a PhD in Telecommunications Engineering from the Universitat Polit\`{e}cnica de Catalunya (UPC) (Barcelona, Spain) in 2014. He is currently a Postdoctoral Researcher in the Network Engineering Department at UPC. His research interests are in management of networks and services for the future Internet. Current focus is on resource management in virtualized networks and functions, software defined networks and cloud computing.
\end{IEEEbiography}

\begin{IEEEbiography}
[{\includegraphics[width=1in,height=1.25in,clip]{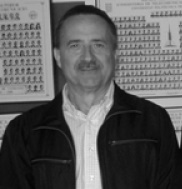}}]{Joan Serrat}
received a degree of telecommunication engineering in 1977 and a PhD in the same field in 1983, both from UPC. Currently, he is a full professor at UPC where he has been involved in several collaborative projects with different European research groups, both through bilateral agreements or through participation in European funded projects. His topics of interest are in the field of autonomic networking and service and network management. He is the contact point of the TM Forum at UPC.
\end{IEEEbiography}

\begin{IEEEbiography}
[{\includegraphics[width=1in,height=1.25in,clip]{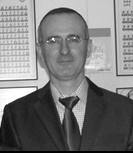}}]{Juan-Luis Gorricho}
received a telecommunication engineering degree in 1993, and a Ph.D. degree in 1998, both of them from the Technical University of Catalonia (UPC). Since 1994 he joined the Department of Network Engineering at the UPC as an assistant professor, and as associate professor since 2001. His most recent research interests have been focused on applying artificial intelligence to the research fields of ubiquitous computing and network management; with special interest on using smart-phones to achieve the recognition of user activities and locations; and applying linear programming and reinforcement learning to solve the network embedding problem and the resource allocation problem, targeting the implementation of the network virtualization and the future network function virtualization.
\end{IEEEbiography}

\begin{IEEEbiography}
[{\includegraphics[width=1in,height=1.25in,clip,keepaspectratio]{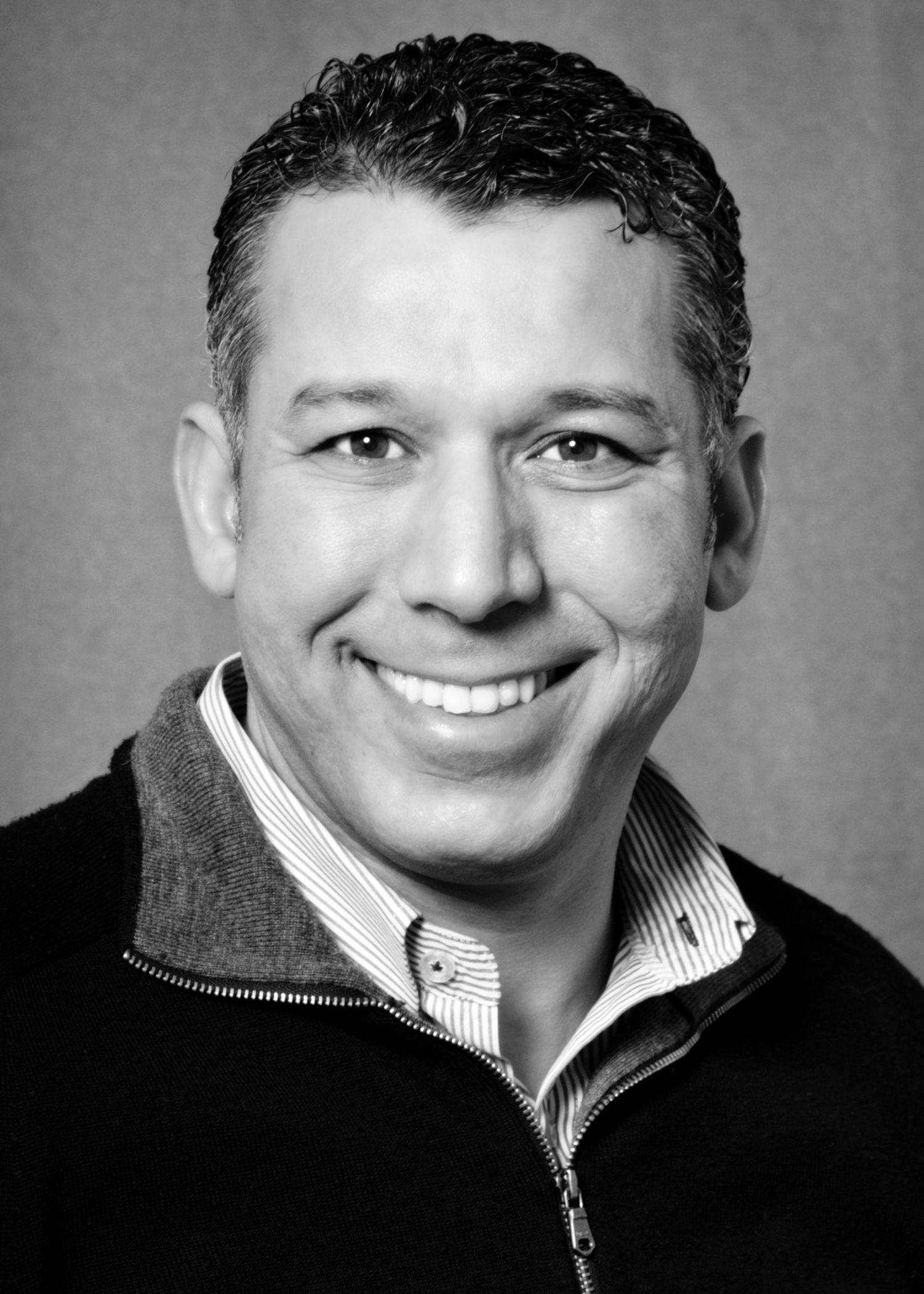}}]{Raouf Boutaba}
received the MSc and PhD degrees in computer science from the Universit\'{e} de Pierre et Marie Curie, Paris, France, in 1990 and 1994, respectively. He is currently a full professor of computer science at the University of Waterloo, Waterloo, ON, Canada, and a distinguished visiting professor at the Pohang University of Science and Technology (POSTECH), Korea. His research interests include network, resource and service management in wired and wireless networks. He has received several best paper awards and other recognitions such as the Premier's Research Excellence Award, the IEEE Hal Sobol Award in 2007, the Fred W. Ellersick Prize in 2008, the Joe LociCero and the Dan Stokesbury awards in 2009, and the Salah Aidarous Award in 2012. He is a fellow of the IEEE and the Engineering Institute of Canada.
\end{IEEEbiography}

\end{document}